\documentclass{article}

\usepackage[english]{babel}
\usepackage{csquotes}
\usepackage[T1]{fontenc}
\usepackage[utf8]{inputenc}
\usepackage{mathtools}
\usepackage{amsmath,amssymb}
\usepackage[hyphens]{url}
\usepackage{graphicx}
\usepackage{txfonts}
\usepackage{lipsum}
\usepackage{subcaption}
\usepackage{lscape}
\usepackage{placeins}           
\usepackage[colorlinks=true,
            linkcolor=black,
            linkbordercolor={0 0 0},
            citecolor=blue]{hyperref}

\usepackage{geometry}
\usepackage{authblk}

\usepackage{caption}
\usepackage{biblatex}
\begin{document}
\title{Satellite-borne $\gamma$-ray astrophysics from coherent interactions in oriented crystals}

\author[1,2]{P. Monti-Guarnieri}
\author[3]{G. Paternò}
\author[3]{A. Sytov}
\author[4]{E. Cavazzuti}
\author[ ]{L. Costamante}
\author[5]{S. Cutini}
\author[5]{M. Duranti}
\author[3,6]{P. Fedeli}
\author[7]{R. J. Gaitskell}
\author[3,6]{V. Guidi}
\author[ ]{V. Haurylavets}
\author[7,8]{S. M. Koushiappas}
\author[1,2]{F. Longo}
\author[9,10]{S. Mangiacavalli}
\author[3,6]{A. Mazzolari}
\author[9,10]{M. Prest}
\author[3,6]{M. Romagnoni}
\author[9,10]{A. Selmi}
\author[11]{M. Soldani}
\author[ ]{V. Tikhomirov}
\author[4]{V. Vagelli}
\author[10]{E. Vallazza}
\author[3]{L. Bandiera\footnote{Corresponding author: bandiera@fe.infn.it}}

\affil[1]{Università degli Studi di Trieste, Dipartimento di Fisica, Via Alfonso Valerio 13, I-34127 Trieste, Italy}
\affil[2]{Istituto Nazionale di Fisica Nucleare (INFN), Sezione di Trieste, Via Alfonso Valerio 13, I-34127 Trieste, Italy}
\affil[3]{INFN, Sezione di Ferrara, Via Giuseppe Saragat 1c, I-44122 Ferrara, Italy}
\affil[4]{Agenzia Spaziale Italiana (ASI), Via del Politecnico snc, I-00133 Roma, Italy}
\affil[5]{INFN, Sezione di Perugia, Via Alessandro Pascoli 23c, I-06123 Perugia, Italy}
\affil[6]{Università degli Studi di Ferrara, Dipartimento di Fisica e Scienze della Terra, Via Giuseppe Saragat 1, I-44122 Ferrara, Italy}
\affil[7]{Department of Physics, Brown University, Providence, RI 02912-9037, USA}
\affil[8]{Brown Center for Theoretical Physics \& Innovation, Brown University, Providence, RI 02912-9037, USA}
\affil[9]{Università degli Studi dell'Insubria, Dipartimento di Scienza e Alta Tecnologia, Via Valleggio 11, I-22100, Como, Italy}
\affil[10]{INFN, Sezione di Milano Bicocca, Piazza della Scienza 3, I-20126 Milano, Italy}
\affil[11]{CERN, Esplanade des Particules 1, 1217 Meyrin, Switzerland}
\renewcommand\Affilfont{\itshape\small}

\date{}
\maketitle

% ----------------------------------------------------------------------
\begin{abstract}
    High-density and high-Z crystals are key elements of most space-borne $\gamma$-ray telescopes operating at gigaelectronvolt energies (such as Fermi-LAT). The lattice structure is usually neglected in the development of a crystalline detector, although its effects on the energy deposit development should be taken into account, since the interactions of a high-energy ($\sim$ 10~GeV) photon or e$^\pm$ impinging along the axis of an oriented crystal are different than those observed in a fully isotropic medium. Specifically, if the angle between a photon (e$^\pm$) trajectory, and the crystal axis is smaller than $\sim$ 0.1$^\circ$, a large enhancement of the pair production (bremsstrahlung) cross section is observed. Consequently, a photon-induced shower inside an oriented crystal develops within a much more compact region than in an amorphous medium. Moreover, for photon energies above a few gigaelectronvolt and incidence angles up to several degrees, the pair production cross section exhibits a pronounced dependence on the angle between the crystal axis and the photon polarization vector. \\

    In this work we show that these effects could be exploited to develop a novel class of light-weight pointing space-borne $\gamma$-ray telescopes, capable of achieving an improved sensitivity and resolution, thanks to a better shower containment in a smaller volume, with respect to non-oriented crystalline detectors. We also show that an oriented tracker-converter system could be used to measure the polarization of a $\gamma$-ray source above few gigaelectronvolts, in a regime that remains unexplorable through any other detection technique. \\

    This novel detector concept could open new pathways in the study of the physics of extreme astrophysical environments and potentially improve the detector sensitivity for indirect dark matter searches in space
\end{abstract}

%\keywords{instrumentation: detectors -- 
%          telescopes -- 
%          relativistic processes -- 
%          polarization -- 
%          techniques: polarimetric}

% ----------------------------------------------------------------------
\section{Introduction} \label{sec_intro_crystals}
Since the 1950s it has been known that the lattice structure can influence the development of electromagnetic (e.m.) processes in oriented crystals. For instance, the alignment of a beam of high-energy e$^\pm$ or photons with crystal axes or planes results in a large increase in the bremsstrahlung and pair production cross sections, respectively (\cite{diambrinipalazzi1962,baryshevski1989,baryshevski1985,baier1998_book,uggerhoj2005,sorensen1987_nature}). When the e$^\pm$ trajectory is nearly parallel to a crystal plane (axis), the successive collisions with the atoms in the same plane (row) are correlated. In this case, one can replace the Coulomb potential and electric field of separate atoms with an average continuous potential. For ultra-relativistic particles, the electric field is Lorentz-boosted by a $\gamma$ factor: it can reach the Schwinger critical field of quantum electrodynamics, the threshold above which nonlinear effects occur in vacuum ($E_0 = m^2 c^3 / e\hbar \sim 1.3 \cdot 10^{16}$~V/cm). This condition is called the strong field (SF) regime and is otherwise reachable only in extreme environments, such as a magnetar atmosphere. 

The SF regime condition is characterized by intense hard-photon emission by e$^\pm$ (\cite{baryshevski1989}), as well as intense pair production by multi-gigaelectronvolts photons (\cite{baryshevski1985}). As a consequence, the e.m. shower initiated by high-energy ($E \gtrsim 10$~GeV) $\gamma$-rays develops in a more compact space. The SF regime effects peak when the angular separation between the incident particle trajectory and the crystal axis is smaller than a characteristic angle ($\Theta_0$), which is nearly independent of the particle energy itself (\cite{baier1998_book, baryshevski1989}).
\begin{equation}
    \Theta_0 = \frac{U_0}{mc^2}
\end{equation}
Here, $U_0$ is the amplitude of the crystal potential well-depth in the laboratory frame, while $mc^2 = 511$ keV is the electron rest-mass energy. In high-$Z$ materials such as tungsten (commonly used as a converter), or high-$Z$ scintillating crystals (employed in homogeneous electromagnetic calorimeters), the critical angle is on the order of a few milliradians. In particular, in W it is approximately $\sim$~1.74 mrad for the $\langle 111 \rangle$ axis (\cite{soldani2023}), and in lead tungstate (PbWO$_4$) $\sim$~0.82 mrad along the $\langle 100 \rangle$ axis (\cite{soldani2025}). For incidence angles exceeding $\Theta_0$, and up to about $1^\circ$, a weaker but still noticeable enhancement of the bremsstrahlung and pair production cross sections remains visible (\cite{soldani2023,soldani2025}).

The SF regime has been experimentally investigated at the CERN SPS extracted beamlines in the 1980s, first by the NA43 and NA48 collaborations (\cite{belkacem1987,kirsebom1998,moore1996}). Pure single-element crystals such as Si, Ge and W were mainly investigated, while the first studies with an oriented scintillator crystal arrived several years later (\cite{baskov1999,bandiera2018,bandiera2019}). Only in the last few years the INFN STORM-OREO collaboration demonstrated that the acceleration of the e.m. shower development results in an increased number of secondary charged particles crossing the volume of an oriented scintillator in full strong field regime, thus increasing the energy deposited by the shower per unit of target thickness (\cite{soldani2025}). This study opened up the possibility of using oriented scintillator crystals to develop oriented calorimeters for applications in particle and astroparticle physics.

For this work we explored the potential of a detector composed by oriented crystals, specifically in the context of high-energy satellite-based $\gamma$-ray astronomy. This paper is structured as follows. In section~\ref{sec_gamma_challenges} we review the limitations of the state of the art $\gamma$-ray detectors, while in section~\ref{sec_oriented_sat} we show how the use of oriented crystals could help in overcoming these limitations. In section~\ref{sec_polarization} we discuss whether and how an oriented tracker-converter system could be sensitive to the $\gamma$-ray polarization above few gigaelectronvolts, a regime otherwise unreachable with conventional detection systems. In section~\ref{sec_gamma_challenges} we report the technological challenges related to the use of a detector composed of oriented crystals in space, and in section~\ref{sec_conclusions} we summarize our findings.

\section{Challenges in the direct detection of high-energy gamma rays} \label{sec_gamma_challenges}
Gamma ray observations give us a direct view into the most extreme environments of the Universe (see e.g., \cite{funk2015,hoffman1999,aharonian2004}). They are one of the most direct ways to study astronomical particle accelerators such as microquasars, supernova remnants, pulsars, active galactic nuclei, and gamma-ray bursts (GRBs). They also provide a unique tool for performing multiwavelength and multimessenger studies, aimed at understanding the origin of cosmic rays and gravitational waves (\cite{funk2015}) and to probe the existence of dark matter (\cite{bringmann2012,leane2019}).

The gigaelectronvolt $\gamma$-ray sky is usually studied by means of space-borne telescopes, such as the Large Area Telescope (LAT) onboard of the Fermi mission (\cite{atwood2009}), or Imaging Air Cherenkov Telescopes (IACTs), although the latter are more suited for energies above several tens of gigaelectronvolts (\cite{gueta2021}). The LAT is a pair conversion detector composed of 16 identical towers: the incident $\gamma$-rays first cross a silicon-tungsten tracker (TKR), where they convert into an electron-positron pair and initiate an e.m. shower. The silicon strip detectors (SSDs) in the TKR measure the e$^\pm$ tracks and thus reconstruct the incidence direction of the parent $\gamma$-ray. Then, all the shower particles are absorbed inside an 8.6~X$_0$ thick imaging calorimeter (CAL) composed of crystal scintillators (CsI(Tl)). A tessellated plastic scintillator anti-coincidence detector (ACD) is placed around the TKR, to filter out the charged cosmic-ray events. The overall volume of the LAT instrument is about 10.1~X$_0$ in depth and $1.3 \times 1.3$~m$^2$ in transverse area. 

In terms of performance, the LAT energy resolution is limited to $\gtrsim$ 10\% at GeV energies, due to the shallowness of the CAL, which prevents a full containment of the e.m. showers. On the other hand, the point spread function (PSF) is limited at low energies by the multiple scattering of the e$^\pm$ through the tungsten plates, and at high energies by the SSDs strip pitch and the plane-to-plane spacing (\cite{ackermann2012}). The peak resolution is $\sim 0.1^\circ$ above 10~GeV. Notably, the LAT performance depends not only on the incident photon energy, but also on the depth at which the conversion occurs (\cite{atwood2009}). In fact, the TKR is divided in two sectors: a front part composed of 12 thin tungsten layers (0.03~X$_0$ each), followed by a back part consisting of four thicker converters (0.18~X$_0$ each). The energy resolution and conversion efficiency of each sector is almost equal, while the PSF for $\gamma$-rays converting in the back sector is worse than the front by a factor of 2-3, due to the increased thickness of the tungsten layers and thus to the stronger impact of the multiple Coulomb scattering (MCS) on event reconstruction. This gap in the PSF is visible for energies up from hundreds of megaelectronvolts to several hundreds of gigaelectronvolts (\cite{atwood2013}).

Since its launch in 2008, Fermi-LAT has made significant strides in $\gamma$-ray detection, unveiling more than 7,000 point sources and almost 100 extended ones (\cite{fermilatDR3,ballet2024}), even if a substantial 30\% of the point {sources} remains unassociated. Further studies, particularly of the extended sources, whose actual extent and physical properties remain under debate (\cite{abdollahi2024}), are therefore required.

Beyond this issue, the scientific community is intrigued by the enhanced $\gamma$-ray emissions detected at the center of the Milky Way, which peaks around a few gigaelectronvolts (\cite{bartels2016,leane2019}). The critical question remains whether this is due to conventional astrophysical sources (\cite{Dinsmore:2021nip}) or to the annihilation of dark matter particles. The energy scale of this excess is also still under discussion: a recent study based on fifteen years of Fermi-LAT data has found a statistically significant halo-like gamma ray excess in the center of the Milky Way, with a spectral peak around 20 GeV and nonnegligible emission from 2 to 200~GeV. Such excess is compatible with the signal produced by the annihilation of a dark matter particle with a mass in the hundreds of gigaelectronvolts (\cite{totani2025}). To solve this puzzle, improved energy, spectral and angular range, and angular resolution are necessary (\cite{charles2014}), alongside a large detector area to reach sufficient statistics with such a relatively low photon flux. Polarization {measurement} at the gigaelectronvolt scale would also significantly help addressing all these problems and clarifying the acceleration processes at the root of the $\gamma$-ray emission. The importance of this kind of measurements was already demonstrated in the last few years by the IXPE mission, sensitive to the photon polarization in the X-ray band (\cite{soffitta2024}). Despite polarimetry being well established in the X-ray and low-MeV range (\cite{ilie2019}), the same technology can't be easily applied to the detection of gigaelectronvolt photons. In fact, no GeV-polarization-sensitive instrument has been developed yet (see section~\ref{sec_polarization}).

At the time of writing, following the recent conclusion of the AGILE mission (\cite{ATel2024_16450}), Fermi-LAT remains the only operational instrument surveying the entire $\gamma$-ray sky at MeV-GeV energies. Space-borne prompt detection of transient high-energy electromagnetic signals is highly {desired} by the community to provide relevant information on the electromagnetic counterparts of gravitational wave events detected on the ground. Thus, several future missions have been proposed to explore the megaelectronvolt band (\cite{fleischhack2021,deangelis2021}); however, no plans currently exist to surpass Fermi-LAT in the gigaelectronvolt regime -- not even with the upcoming Imaging Air Cherenkov Telescopes (IACTs), which, despite their much larger effective areas, generally offer poorer angular and energy resolution below a few hundred gigaelectronvolts (\cite{gueta2021}). In fact, there are no concrete plans for high-energy $\gamma$-ray observatories once Fermi ceases operations. Multipurpose cosmic-ray missions such as the already operational DAMPE and CALET, or the forthcoming HERD experiment, could partially compensate for this gap, thanks to their deep calorimeters ($\sim$~50~X$_0$). Nonetheless, their limited angular resolution, restrictive trigger configurations, and orbital parameters, will significantly constrain their overall $\gamma$-ray detection performance (\cite{kai2025,kyratzis2022}).

In conclusions, it is time to develop a novel class of $\gamma$-ray instruments, capable of achieving better performance than Fermi-LAT in terms of effective area, acceptance, energy and angular resolution, and especially cost/performance ratio. A detector composed of oriented crystals and employed in pointing strategy could be the answer to some of these challenges and could help overcoming the limitations of direct gigaelectronvolt $\gamma$-ray observations, potentially revolutionizing the field in the post-Fermi era.

Seminal ideas on exploiting oriented crystals to enhance gamma ray detection in space-borne telescopes were proposed decades ago (\cite{moller1989}), but they were never pursued in detail. Recent results from the INFN STORM-OREO collaboration have revived this direction: in particular, the development of end-to-end Monte Carlo models, validated against dedicated experimental measurements, now enables more realistic and quantitatively solid design studies and proposals.

\section{A space-borne $\gamma$-ray telescope based on oriented crystals} \label{sec_oriented_sat}

In this section we discuss how the use of oriented crystals in the development of a {Fermi-LAT-like} tracker and calorimeter could improve their performance, for all $\gamma$-rays incident within a fraction of a degree from the crystal axes. For larger incidence angles, the detector would perform exactly as a non-oriented one, thus the use of oriented crystals would not damage the overall performance of the detector. 

We discuss the case of an oriented silicon-tungsten tracker and of an oriented calorimeter in two separate sections. For the calorimeter, we considered the use of oriented PbWO$_4$ crystals instead of the more common CsI(Tl), since it has been demonstrated that the former can be easily grown oriented (\cite{bandiera2018,soldani2023,soldani2025}) or cut along a crystalline plane after growth, while the same is not true in general for all crystalline materials, such as CsI(Tl). Moreover, the PbWO$_4$ high Z, high density, and good radiation tolerance make it particularly suited for the development of this kind of detectors. For instance, PbWO$_4$ was already used for the development of the CALET electromagnetic calorimeter (\cite{torii2007}).

The results presented in this work were obtained by means of the Geant4 version 11.2 toolkit (\cite{agostinelli2003,allison2016}). The Physics List used in this case is a modified version of the \texttt{FTFP\_BERT} list, which implements the standard high-energy e.m. and hadronic processes, including also photo-nuclear interactions (\cite{allison2016}). The modification was necessary, since until very recently Geant4 did not implement the physics of oriented crystals: a library dedicated to the full simulation of interactions in oriented tungsten crystals was released only in Geant4 version 11.2, implementing the Monte Carlo methods described in (\cite{bandiera2015,sytov2019,sytov2023,negrello2025novel}). Thus, we used that version for the simulation of interactions in an oriented W crystal, while for PbWO$_4$ a custom version was still required. We plan to enable the possibility to simulate PbWO$_4$ crystals, adopting both the simplified method described here and the full simulation approach, in a future release of Geant4.

In the modified version used for this work, the differential cross sections of the bremsstrahlung and pair production processes were multiplied by a set of coefficients, which increase with the particle energy (\cite{bandiera2019,baryshevsky2017}). These factors were computed beforehand through a full Monte Carlo simulation, where the electromagnetic interaction probabilities in the axial field of a W or PbWO$_4$ lattices were computed by directly integrating the quasi-classical Baier-Katkov formula on realistic particle trajectories (\cite{guidi2012,bandiera2015,sytov2019,sytov2023}). The dependence of the cross sections on the relative angle between the particle trajectories and the crystal orientation is taken into account as well. This approach has been extensively validated in several studies performed at the CERN PS and SPS by the STORM-OREO collaboration employing either W or PbWO$_4$ crystals, {exposed to beams of electrons, positrons, photons, muons and charged hadrons in the $1-200$~GeV energy range} (\cite{bandiera2018,bandiera2022,soldani2023,soldani2025}). Thus, it can be considered as a reliable way to simulate the interactions occurring in an oriented crystal exposed to a high-energy particle beam.

\subsection{An oriented silicon-tungsten tracker} \label{sec_oriented_tkr}
\begin{figure}[ht!]
    \centering
    \resizebox{\hsize}{!}{\includegraphics{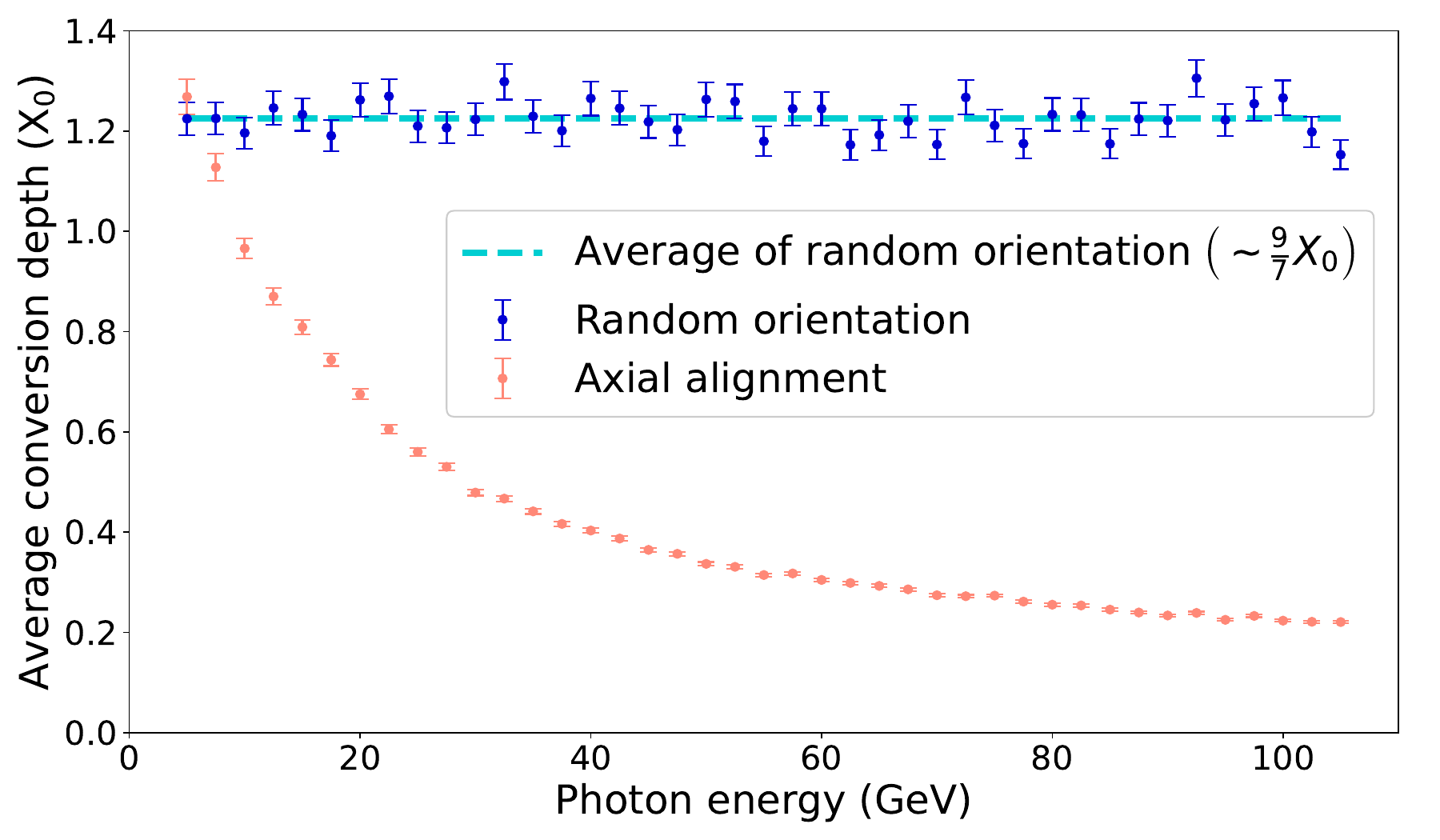}}
    \caption{Average depth at which high-energy photons convert in an {$\text{e}^+ - \text{e}^-$} pair, inside an infinitely thick and infinitely large tungsten crystal aligned along the $\langle 111 \rangle$ axis. The blue dots (red dots) correspond to the case of photons incident on the non-aligned (perfectly aligned on-axis) crystal. Each point shows the average value over 10,000 simulations.}
    \label{fig:Wcrys_PPenhancement}
\end{figure}

\begin{figure}[ht!]
    \centering
    \resizebox{\hsize}{!}{\includegraphics{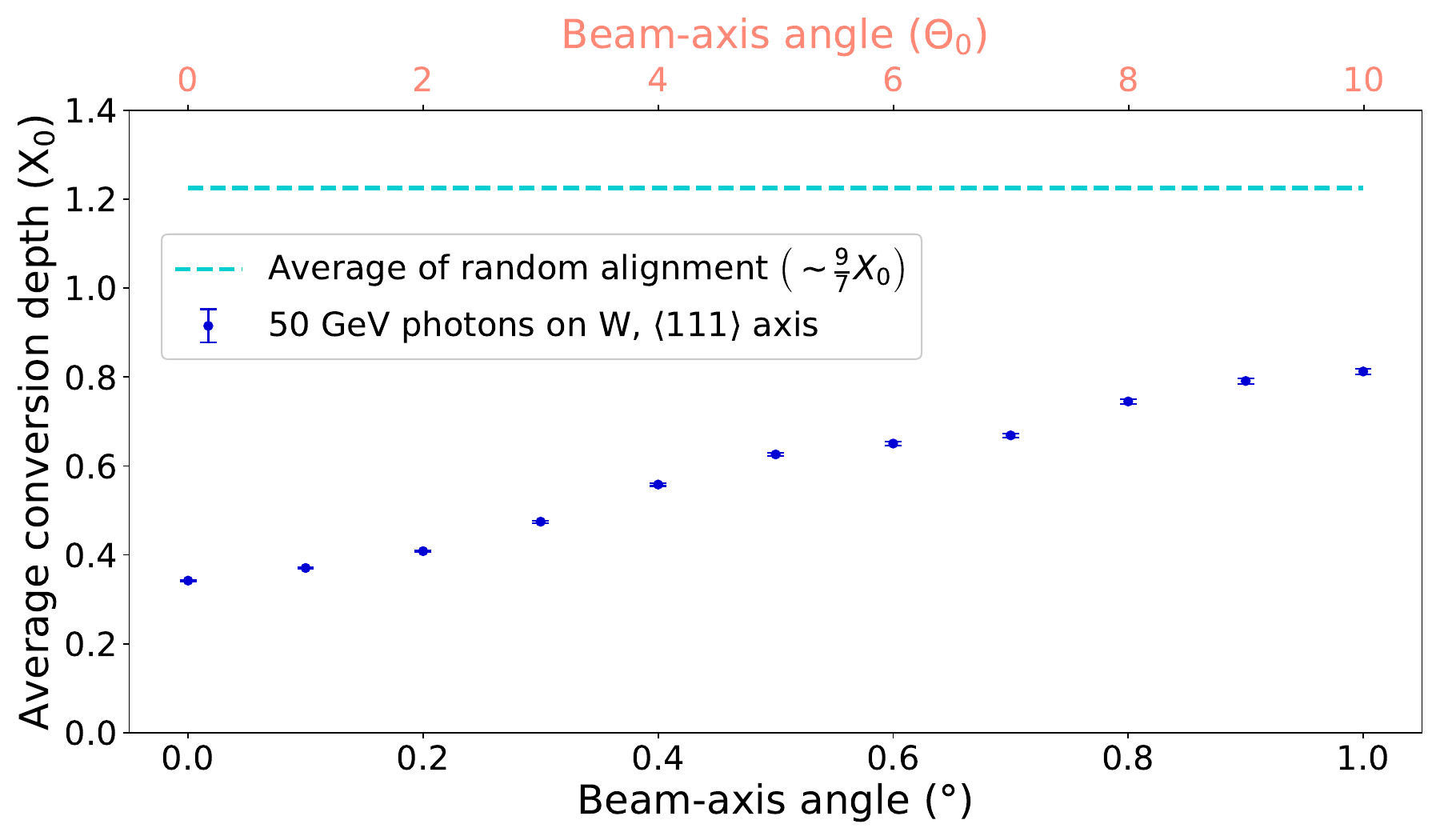}}
    \caption{Average depth at which a 50~GeV photon converts in an {$\text{e}^+ - \text{e}^-$} pair, inside an infinitely thick and infinitely large W crystal, aligned along the $\langle 111 \rangle$ axis. Different beam-axis angles are shown. The upper horizontal axis measures the angle in units of the SF critical value ($\Theta_0$), which is $\sim$~1.74~mrad ($\sim 0.1^\circ$) for the W $\langle 111 \rangle$ axis (\cite{soldani2023phd}). Each point shows the average value over 10,000 simulations.}
    \label{fig:Wcrys_PPenhancement_anglescan}
\end{figure}

Let us consider now the case of a Fermi-LAT-like silicon-tungsten tracker. We see that the tungsten layers could be fabricated as crystalline (\cite{moore1996,kirsebom1998,chehab2002,artru2005,bandiera2022,soldani2023}): in case of a perfect orientation, this would allow the high-energy photons to convert much earlier, as shown in figure~\ref{fig:Wcrys_PPenhancement}. This effect peaks if the incidence angle is smaller than the SF critical angle ($\Theta_0$), but it can be seen that even at relatively large angles ($\sim 1^\circ$) the average depth of the first conversion is strongly reduced (figure~\ref{fig:Wcrys_PPenhancement_anglescan}). This is much different from the standard Bethe--Heitler pair production, where the photon conversion length is $\sim (9/7)$~X$_0$ at all energies (\cite{fabjan2003}).

The higher conversion probability opens the door to two equally interesting possibilities. The first is to develop a tracker with a thickness comparable to that of the Fermi-LAT, but incorporating crystal orientation -- effectively plugging in the orientation into a conventional, non-oriented detector design. In such a configuration, {most on-axis high-energy photons would convert much earlier than normal, at depths comparable to the thickness of the TKR front sector. In the LAT, the PSF for front-converting events is better than the back-converting by a factor of two or three (\cite{atwood2009}), due to the reduced MCS of the e$^\pm$ before passing through the silicon detectors. Thus, an oriented detector with a Fermi-LAT-like structure could achieve the front PSF for all on-axis high-energy gamma rays, not only for the front-converting ones. In practice, this would lead to an average improvement of the PSF by a factor of $2-3$, visible from around 10~GeV (the SF scale) up to hundreds of gigaelectronvolts. At even higher energies, the PSF would still be limited by the tracker plane-to-plane spacing and the SSD strip pitch.} In summary, such a detector would behave conventionally for off-axis particles, while providing a reduced field of view (FOV) but enhanced performance for on-axis events.

The second possibility is to develop a thinner tracker than that of the Fermi-LAT, to achieve a photon conversion efficiency and PSF comparable to the LAT, but only for on-axis high-energy events (i.e., in the SF regime). In this scenario, the smaller material budget would result in a reduced detection efficiency for off-axis photons, as well as for all incident low-energy $\gamma$-rays. However, the reduced thickness would also mitigate the effects of multiple scattering on event reconstruction, since its angular spread scales approximately with the square root of the tracker thickness (\cite{charles2014,fabjan2003}). Thus, the off-axis PSF would improve in the energy range where MCS is the most important contribution to the angular resolution, which is approximately up to $\sim$~3~GeV (\cite{atwood2009}).

To provide a more quantitative estimate of the trade-offs of this second scenario, we selected a few photon energies of interest (E$_\gamma$). We then considered TKR-like detectors with thickness ($t$) given by the depth at which, on average, an on-axis photon with energy $E_\gamma$ converts inside an oriented tungsten crystal (figure~\ref{fig:Wcrys_PPenhancement}). In this way, we are considering detectors which are smaller than the LAT and achieve roughly the same sensitivity and PSF for on-axis $\gamma$-rays with energy $\geq E_\gamma$. We then computed the photon detection efficiency ($\eta$) of these detectors, namely the fraction of off-axis photons converting in a thickness smaller than $t$, assuming that the conversion probability follows an exponential distribution with average value equal to $(9/7) X_0$. Finally, we computed the characteristic MCS angle ($\theta_{rms}$) for an e$^\pm$ propagating through the whole tracker thickness (\cite{charles2014}).
\begin{equation}
    \theta_{rms}(E, t) \approx \frac{\sqrt{2} \cdot 13.6 \, \text{MeV rad}}{E} \sqrt{t} \left[ 1 + 0.038 \log(t) \right]
\end{equation}   
This angle is roughly proportional to the PSF for off-axis photons, at energies $\lesssim$~3~GeV (\cite{atwood2009}). It can be seen that a reduction of the TKR thickness by a factor of $1.5-2$ translates in an improvement of $\theta_{rms}$ and the PSF by $30-60\%$, while causing a proportional loss in detection efficiency (table~\ref{table:sec3_1b}).

In summary, a thinner oriented detector would remain highly sensitive to on-axis $\gamma$-rays up to very high energies and simultaneously provide an improved PSF for low-energy off-axis events, although at the price of a lower detection efficiency. This loss of statistics could be compensated by enlarging the detector effective area, thus requiring no additional cost compared to a non-oriented design, owing to the overall thickness reduction.

It is worth noticing that the PSF improvements achievable in both scenarios could be fundamental not only for a better resolution of crowded regions in the Galactic Plane, but also for the measurement of the $\gamma$-ray polarization above 100~MeV. In fact, the latter requires a precise reconstruction of the distribution of the e$^\pm$ opening angles, with respect to the incidence direction, something that is not usually achievable at low energies precisely due to the MCS (\cite{bernard2024}).

\begin{table}
    \centering
    \begin{tabular}{c c c c}
    \hline\hline
    E$_\gamma$ (GeV) & $t$ (X$_0$) & $\eta$ reduction (\%) & $\theta_{rms}$ reduction (\%) \\
    \hline
        10 & 0.97 & 7.07 & 5.84 \\
        25 & 0.56 & 37.87 & 29.79 \\
        50 & 0.34 & 59.46 & 46.64 \\
    \hline
    \end{tabular}
    \caption{Reduction of the average angular dispersion and detection efficiency due to multiple Coulomb scattering for different tracker thicknesses. Each row represent a different configuration, such that, on average, an on-axis photon with energy higher than $E_\gamma$ (first column) converts at a depth smaller than the tracker thickness (second column). The third column reports the relative reduction in detection efficiency between each configuration and the LAT, for off-axis incident photons. The fourth column shows the corresponding variation in the characteristic MCS angle ($\theta_{rms}$) for off-axis low-energy photons ($\lesssim$~3~GeV), which is proportional to the PSF improvement achieved in the same energy range.}
    \label{table:sec3_1b}
\end{table}

\subsection{An oriented electromagnetic calorimeter} \label{sec_oriented_cal}
\begin{figure}[ht!]
    \centering
    \resizebox{\hsize}{!}{\includegraphics{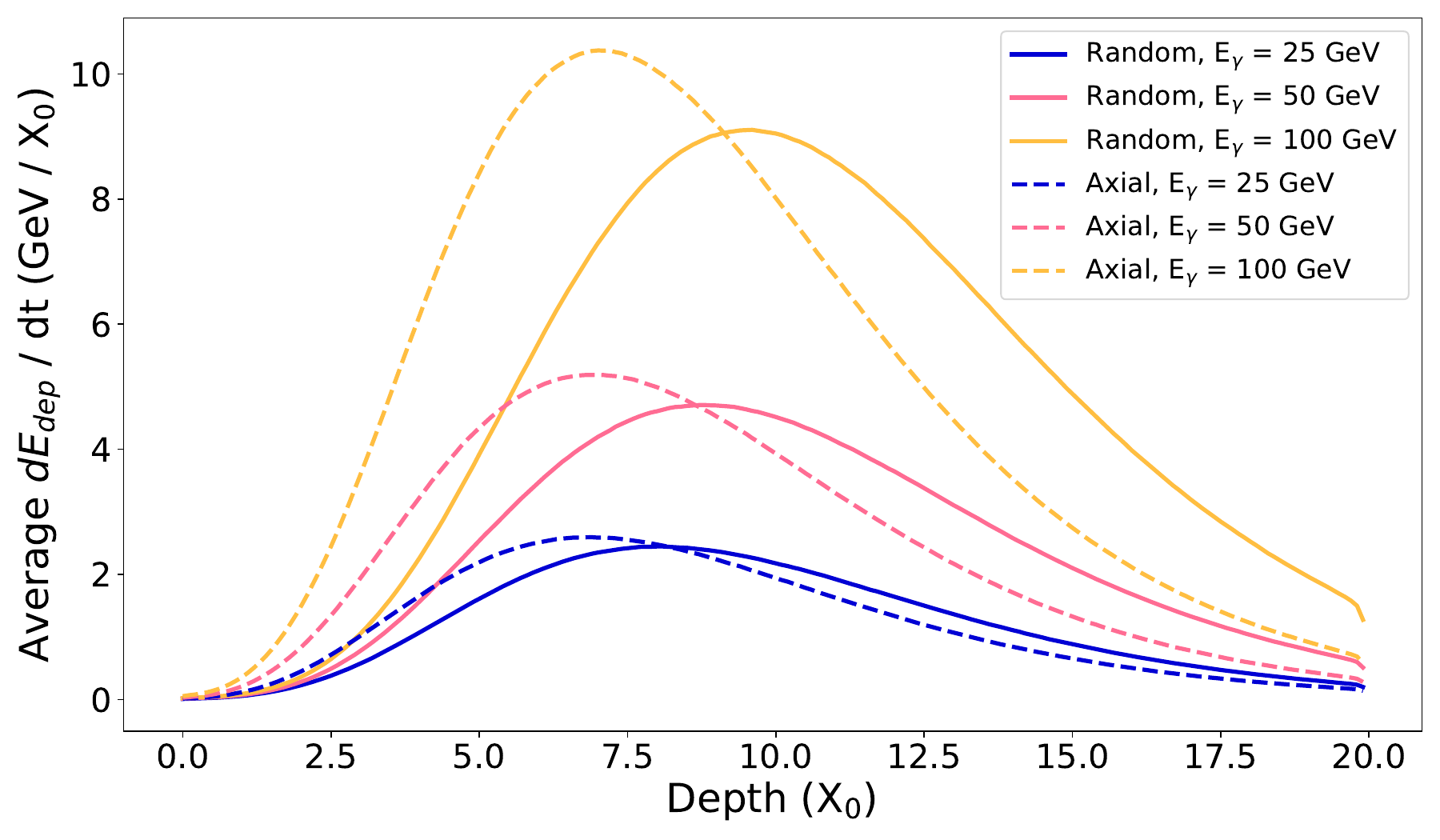}}
    \caption{Average energy deposited by high-energy photons per unit thickness inside an infinitely thick and large PbWO$_4$ crystal aligned along the $\langle 100\rangle$ axis. The solid (dashed) lines represent the case of photons incident on a non-aligned (axially aligned) crystal. Each line is the average value over 10,000 simulations.}
    \label{fig:PWOcrys_Edep_vs_t}
\end{figure}

\begin{figure}[ht!]
    \centering
    \resizebox{\hsize}{!}{\includegraphics{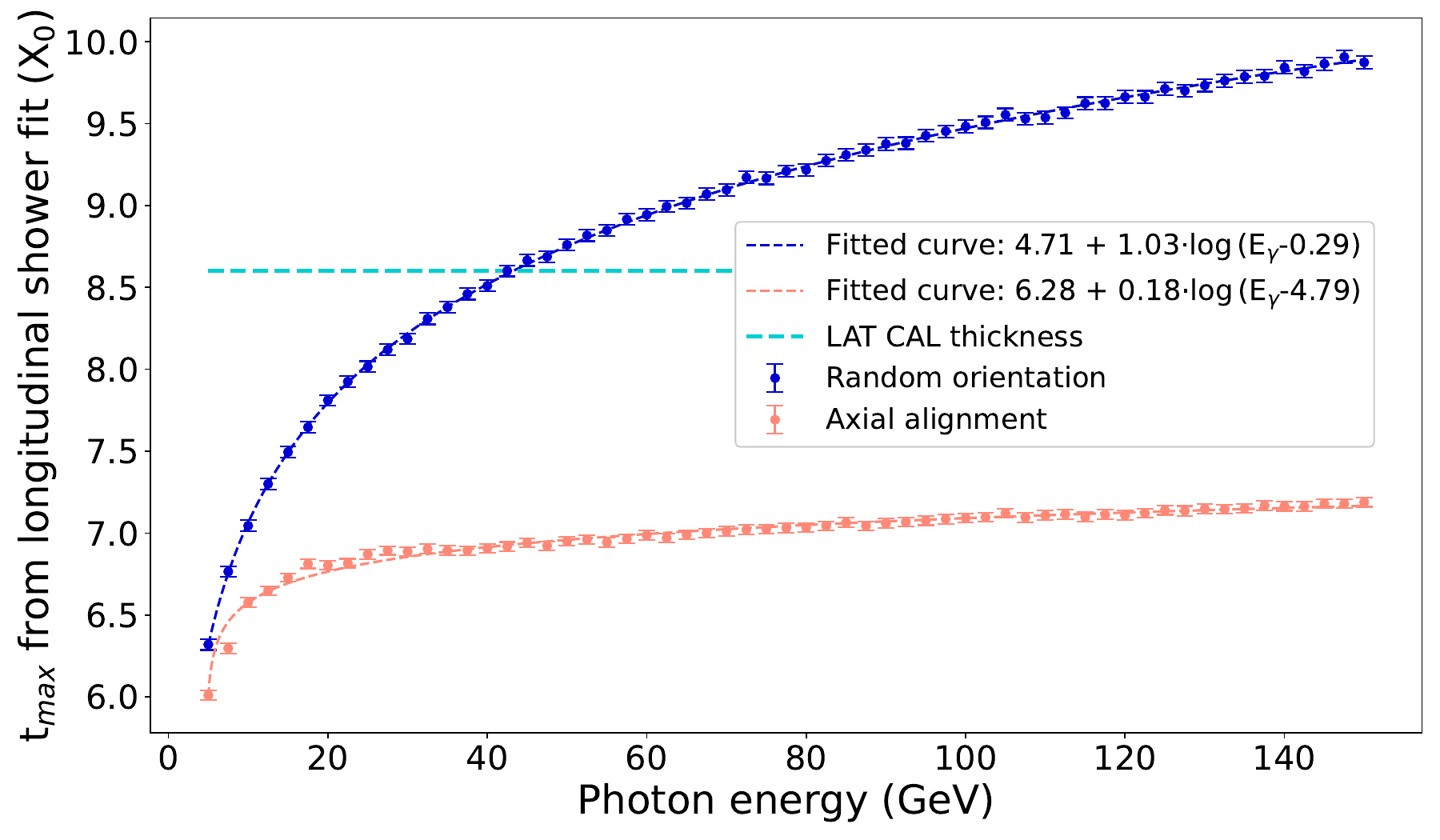}}
    \caption{Variation in the longitudinal depth of the maximum energy deposit ($t_{\max}$), as a function of the $\gamma$-ray energy. The depths were computed by fitting the shower longitudinal profiles with the standard gamma function defined in \cite{fabjan2003}: $dE/dt = c \cdot ( t^{a-1} e^{-bt} )$. The depth of the maximum energy deposit is then $t_{\max} = (a-1)/b$. The blue dots (red dots) represent the case of photons incident on the non-aligned (axially aligned) crystal. The logarithmic fits on the $t_{\max}$ curves were done to show the acceleration of the e.m. shower occurring in an oriented crystal, with respect to an isotropic material. While the logarithmic dependence of the isotropic case is well known (\cite{fabjan2003}), the functional form shown for the oriented case is not reported in any previous study. Each point shows the average value over 10,000 simulations.}
    \label{fig:PWOcrys_tmax_vs_Ein}
\end{figure}

Thanks to the SF acceleration of the e.m. shower, a hodoscopic calorimeter composed of oriented scintillating crystals could contain a significantly higher fraction of the energy of the incident photons, with respect to a calorimeter made of non-aligned crystals, as shown in figure~\ref{fig:PWOcrys_Edep_vs_t}. Since the SF regime effects grow with the energy of the incident particles, the acceleration of the e.m. shower development grows as well. This partially counteracts the well known fact that the depth of the maximum energy deposit in an isotropic material scales as the logarithm of the particle energy (\cite{charles2014}). As a result, inside an oriented crystal, the depth of the maximum energy deposit depends only weakly on the energy of the incident particle (\cite{soldani2023}) and it would remain within the LAT physical thickness for on-axis photons with energies well up to hundreds of gigaelectronvolts (figure~\ref{fig:PWOcrys_tmax_vs_Ein}) or even more. In fact, the SF acceleration of the e.m. shower grows in intensity and saturates only at TeV energies (\cite{uggerhoj2005}).

\begin{figure}[ht!]
    \centering
    \resizebox{\hsize}{!}{\includegraphics{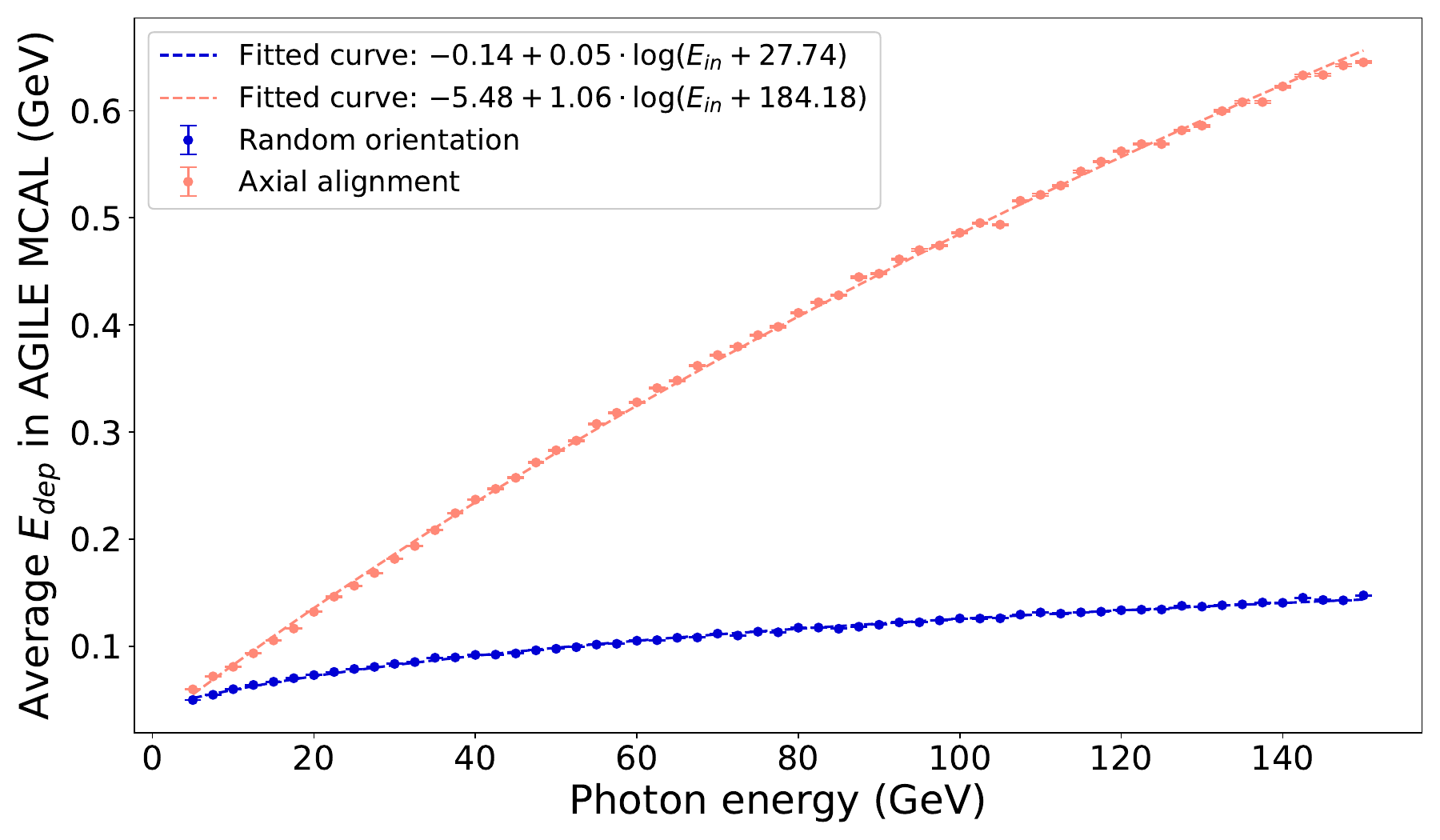}}
    \caption{Average energy deposited by high-energy photons inside an AGILE/MCAL-like calorimeter (1.5~X$_0$, \cite{tavani2009}) composed of oriented PbWO$_4$ crystals, aligned along the $\langle 100\rangle$ axis. The solid (dashed) lines represent the case of photons incident on a non-aligned (axially aligned) crystal. Each point shows the average value over 10,000 simulations.}
    \label{fig:Edep_vs_Ein_AGILE_CALdepth}
\end{figure}

\begin{figure}[ht!]
    \centering
    \resizebox{\hsize}{!}{\includegraphics{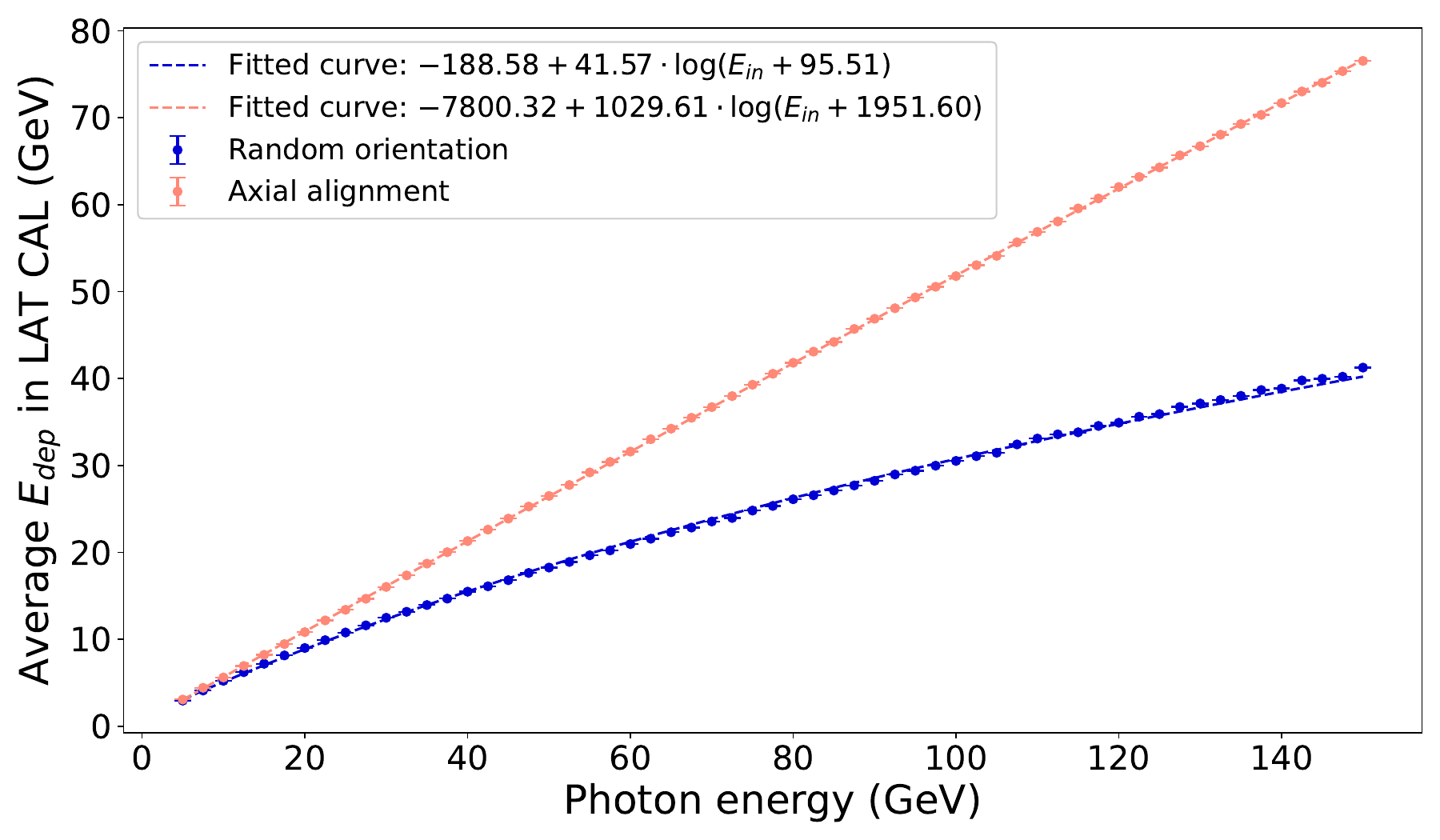}}
    \caption{Average energy deposited by high-energy photons inside a Fermi-LAT-like calorimeter (8.6~X$_0$, \cite{atwood2007}) composed of oriented PbWO$_4$ crystals, aligned along the $\langle 100\rangle$ axis. The solid (dashed) lines represent the case of photons incident on a non-aligned (axially aligned) crystal. Each point shows the average value over 10,000 simulations.}
    \label{fig:Edep_vs_Ein_CALdepth}
\end{figure}

The above discussion suggests that an oriented calorimeter could achieve a significant extension of its energy band-pass and possibly also an improvement of its energy resolution, at no additional cost, especially if the chosen material is easily grown as oriented, as it happens for PbWO$_4$ (\cite{senguttuvan1998}). This is true not only for detectors as thick as the LAT: on the contrary, even greater improvements could be achieved by orienting a smaller detector, such as the AGILE/MCAL (\cite{tavani2009}) (1.5~X$_0$ thick). In fact, as already shown in (\cite{soldani2025}), the energy deposited by an incident on-axis photon in such a thin detector can increase by a factor of 3, with respect to the off-axis ones (figure~\ref{fig:Edep_vs_Ein_AGILE_CALdepth}), while for a Fermi-LAT-like calorimeter the increase factor would be only $\sim 1.5$ (figure~\ref{fig:Edep_vs_Ein_CALdepth}). The corresponding effective increase in the calorimeter thickness for Fermi-LAT would be from $\sim 20\%$ up to $\sim 40\%$, for photons up to hundreds of gigaelectronvolts. These results naturally lead to two possible detector designs, in perfect analogy to what was discussed discussed at the end of subsection~\ref{sec_oriented_tkr}.

The first design is a calorimeter with a thickness comparable to that of Fermi (or AGILE/MCAL), but incorporating crystal orientation, to extend its accessible energy range while improving also its energy resolution. Furthermore, hadron–photon discrimination would be enhanced, as hadronic showers are unaffected by lattice orientation, thereby amplifying the intrinsic differences between hadronic and electromagnetic cascades (\cite{montiguarnieri2024}).
    
The second design is a calorimeter thinner than Fermi-LAT, yet equally sensitive to on-axis high-energy photons while being less responsive to off-axis particles. The reduced thickness could permit a reduction of the weight to be launched in orbit and/or an increase in the effective area, a crucial improvement for compensating the low fluxes of high-energy $\gamma$-rays.

\begin{figure}[ht!]
    \centering
    \resizebox{\hsize}{!}{\includegraphics{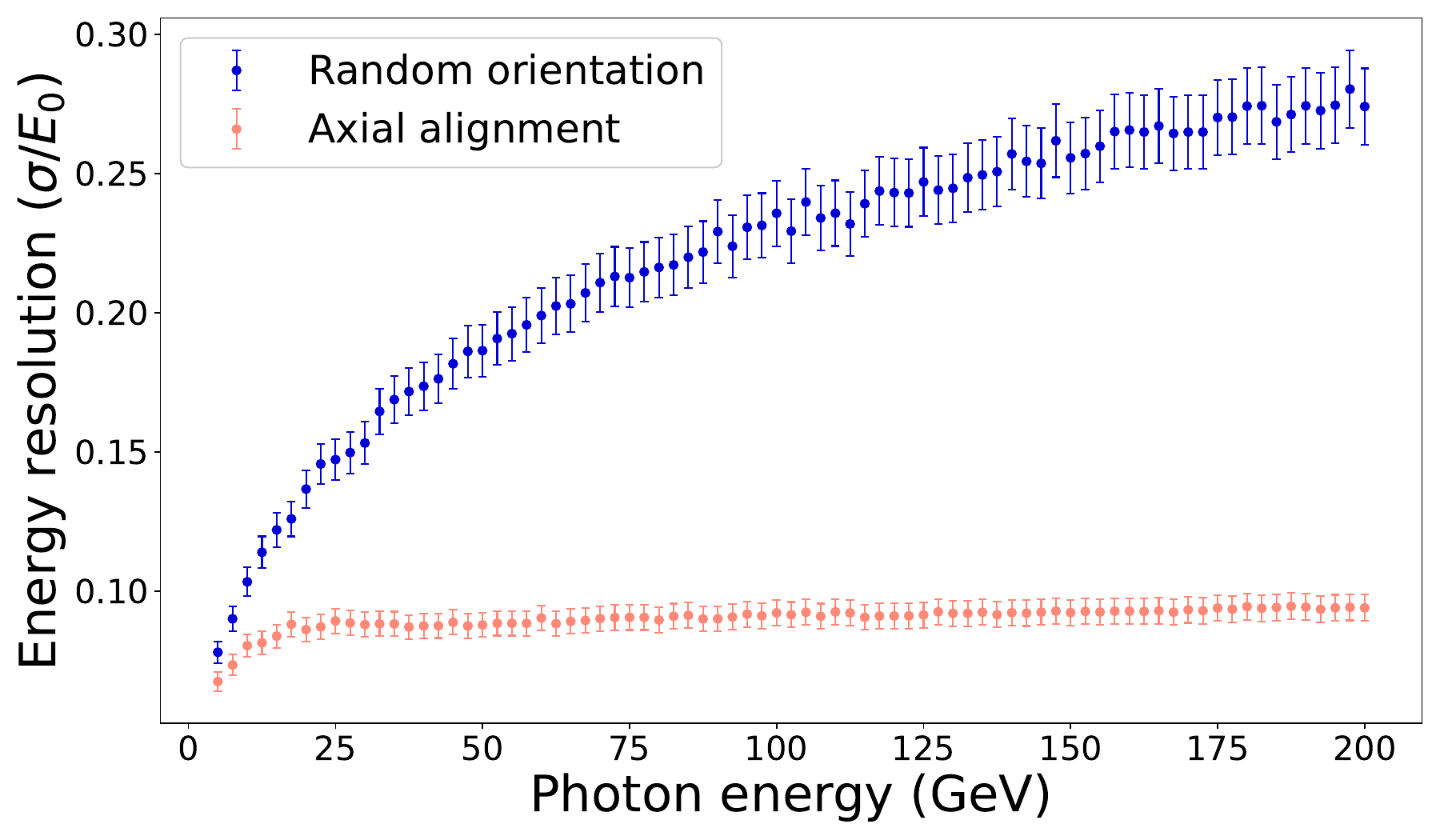}}
    \caption{Energy resolution obtained by fitting the shower development curve for photons incident on a Fermi-LAT-like calorimeter. The resolution was computed as the relative error on the reconstructed photon energy ($\sigma/E_0$). The blue dots (red dots) represent the case of photons incident on the non-aligned (axially aligned) crystal. Each point shows the average value over 10,000 simulations.}
    \label{fig:PWO_reco2_E0res_vs_Ein}
\end{figure}

To assess more quantitatively the energy resolution improvement achievable in this scenario, we have simulated the photon shower development inside a randomly aligned and an oriented calorimeter, with the same longitudinal segmentation as the Fermi-LAT (8 crystal layers, $1.075$~X$_0$ thick). For simplicity, we have not considered any segmentation in the transverse plane. We simulated 600,000 showers for each configuration, with initial photon energies uniformly distributed between 2.5~GeV and 150~GeV. We fitted the energy deposit curve in the calorimeter with the standard gamma parameterization defined in \cite{fabjan2003}.
\begin{equation}
    \frac{dE}{dt} = E_0 b \frac{(b t)^{a-1} e^{-b t}}{\Gamma(a)}
\end{equation}
Here $t$ is the longitudinal depth expressed in units of radiation length, $E_0$ is the incident photon energy and $a$ and $b$ are material-dependent coefficients. Figure~\ref{fig:PWO_reco2_E0res_vs_Ein} shows the variation in the energy resolution, computed as the relative error on $E_0$ (free parameter in the fit) as a function of the incident photon energy. Notably, the resolution measured in the non-aligned case is worse than the actual Fermi-LAT energy resolution, due to the relatively simple and approximated method used in this case. In fact, in the Fermi-LAT event reconstruction, the energy of the incident photon is estimated with a full 3D fit of the shower development in the CAL, and secondary correction factors derived from Monte Carlo simulations are applied to further refine the energy measurement (\cite{atwood2013}). However, we do not think that this discrepancy is an issue, with respect to our results. In fact, the key result we found is that, even with a simple approach to the energy reconstruction, the resolution obtained in the oriented case shows little to no dependence on the incident photon energy above $\sim$ 10~GeV, while it does so in the randomly aligned case, as in the real Fermi-LAT performance curves. This is due to the fact that, in the aligned case, the shower peak remains well contained by the calorimeter volume, as shown already in figure~\ref{fig:PWOcrys_tmax_vs_Ein}: the fit is thus accurately performed even as the energy grows. This means that an oriented Fermi-LAT-like calorimeter could be able to achieve the resolution measured at 10~GeV for all energies above this value. In the specific LAT case, this would mean limiting the energy resolution to $\sim$ 7.5\%, yielding a net improvement of up to $5-15\%$ for hundreds of gigaelectronvolts photons.

A small-scale prototype calorimeter composed of oriented PbWO$_4$ crystals is currently being developed by the OREO collaboration (\cite{bandiera2023, malagutti2024, bandiera2025}). Beamtest campaigns at the CERN PS and SPS have already demonstrated the feasibility of inter-aligning the towers inside an oriented calorimeter with the precision required to guarantee the simultaneous alignment of all crystals at the same time. The results of these campaigns will be discussed in a dedicated paper, currently in preparation.

\newpage
\section{The dream of measuring the photon polarization at the gigaelectronvolt scale} \label{sec_polarization}

Important information on the mechanisms underlying the high-energy photon emission process can be extracted by measuring the $\gamma$-ray polarization. In fact, emission processes such as synchrotron, curvature, {inverse Compton scattering}, and magnetic photon splitting, can give rise to various degrees of linear polarization (\cite{ilie2019}). Despite the inherent challenges related to the polarization measurement, in recent years, some instruments have achieved important milestones in hard X-ray and soft $\gamma$-ray polarimetry: in a sense, with sensitive missions such as IXPE we entered in the X-ray polarimetry era. Soft $\gamma$-ray polarimetry at the multi-kiloelectronvolt and megaelectronvolt scale (i.e., in the Compton regime) is also well established (\cite{delmonte2022}). On the other hand, $\gamma$-ray polarimetry at the gigaelectronvolt scale (i.e., in the full Pair Production regime) is still a terra nullius. In fact, the measurement of the opening angle of an electron-positron pair produced after the conversion of a high-energy photon is essentially prevented by the e$^\pm$ MCS of the particles in the detector. To overcome this limit, some approaches have been proposed in the past, such as the use of gas time-projection chambers (\cite{bernard2013}), emulsion-based detectors (\cite{takahashi2015}), or Fermi-LAT-like trackers entirely made of silicon (\cite{bernard2022}). However, at the time of writing, all these approaches remain limited in terms of energy range and effective area, and none has been developed beyond the level of a small-scale proof-of-concept.

\begin{figure}[ht!]
   \centering
   \resizebox{\hsize}{!}{\includegraphics{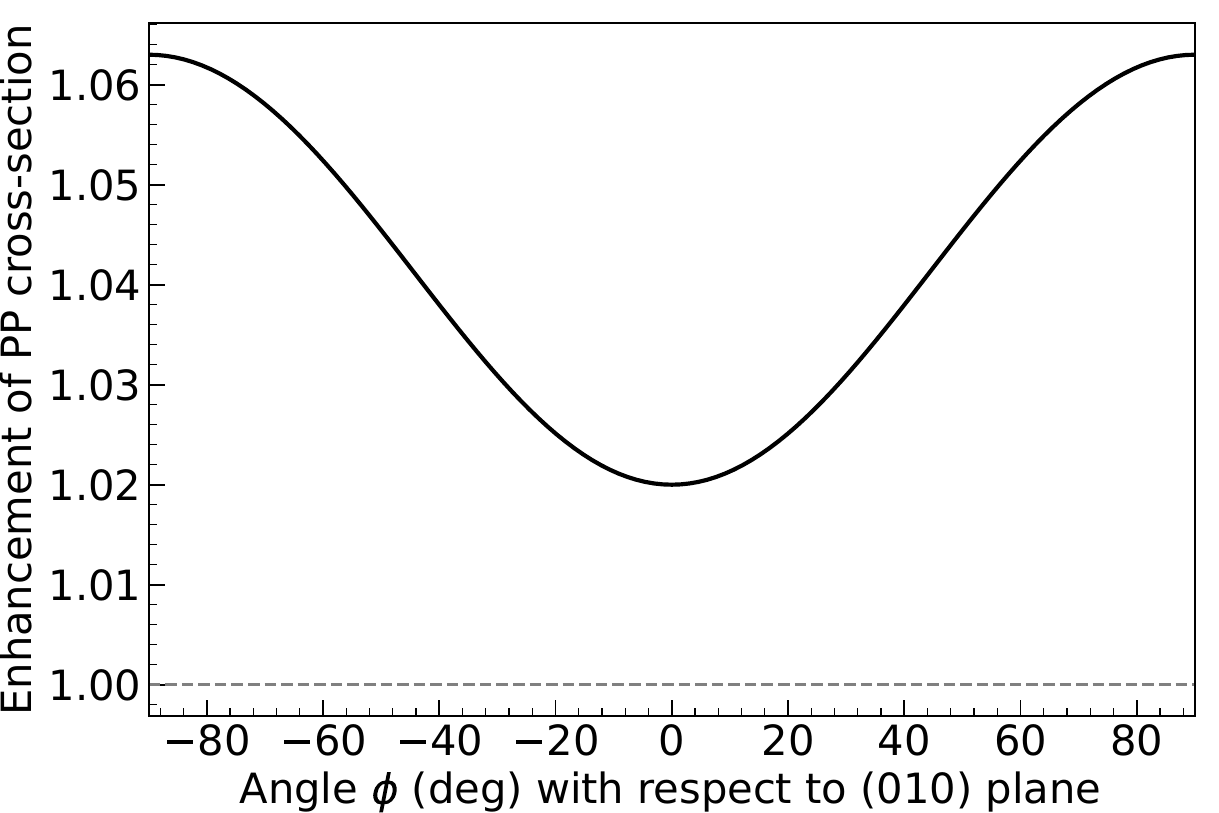}}
   \caption{Pair production cross section enhancement, with respect to the Bethe--Heitler value for an isotropic medium, obtained with a 3~GeV fully linearly polarized $\gamma$-ray beam impinging on an oriented cooled copper radiator at an angle of about $\theta$ = 10 mrad from the $\langle 101 \rangle$ main axis. Here, $\phi$ is the angle of the polarization vector with respect to the (010) plane.}
   \label{fig:polar_xsec}
\end{figure}

An interesting alternative is offered once again by oriented crystals. As it was first noted by Cabibbo and others in the early 1960s (\cite{cabibbo1962}), the pair production cross section for photons moving along an axis or plane of an oriented crystal heavily depends on its linear polarization state (figure~\ref{fig:polar_xsec}). This is much different from the standard Bethe-Heitler pair production, where the photon polarization only affects the angular distribution of the outgoing {$\text{e}^+ - \text{e}^-$} pairs (\cite{cabibbo1962}). It is worth noticing that this is a different regime from the strong field discussed in the previous sections. Following the seminal proposals by Cabibbo and collaborators (\cite{cabibbo1962,cabibbo1962circular,palazzi1968high}), an oriented single crystal can serve as an analyzer of high-energy $\gamma$-ray linear polarization through the coherent pair production (CPP) process.
When the photon formation length becomes comparable to the crystal periodicity, the pair production amplitude interferes constructively at reciprocal lattice vectors. The CPP cross section acquires a dependence on the angle between the photon polarization and a selected crystal plane/axis, leading to an azimuthal modulation of the $e^+e^-$ yield. Operating an analyzer at two orthogonal azimuths ($\parallel,\perp$) one can measure the polarization asymmetry ($A$), which gives the photon linear polarization fraction ($P_\gamma$), once the analyzing power ($\mathcal{A}_{\mathrm{eff}}$) is known, either through an experimental calibration or through numerical calculations starting from the crystal orientation, material, and thickness.
\[
A \equiv \frac{N_{\parallel}-N_{\perp}}{N_{\parallel}+N_{\perp}}
= \mathcal{A}_{\mathrm{eff}}(E_\gamma,\theta)\,P_\gamma
\]
The enhancement of the pair production cross section under CPP has a smaller value than the SF one, but a much lower energy threshold and a much wider angular acceptance (with visible enhancements achievable even at $5^\circ-10^\circ$ from the crystal axis). The observation by Cabibbo and colleagues was later confirmed by an experimental team located in Frascati with few gigaelectronvolt photon beams. A further proof was obtained also by the CERN NA59 collaboration, which used thin Ge and diamond crystals both to produce and detect polarized photon beams in the 10-100~GeV energy band, thus achieving an effective multi-GeV polarimeter (\cite{apyan2008}).

In this sense, a hypothetical polarization-sensitive detector could be made by using an oriented converter made of diamond or copper, materials which provide large analyzing power, followed by multiple silicon tracking detectors. If the converting photon is linearly polarized, the pair production cross section would depend on the misalignment between the photon polarization direction and the crystallographic planes. Thus, to determine the polarization fraction of any source in the gamma ray sky, it would be necessary to point at it and measure the average conversion rate at different source-detector angles. This could be achieved either by rotating the instrument spacecraft from time to time, or by deploying two identical light-weight detectors which observe the same source at the same time, with different alignments. The second option would be more expensive, but it would reduce the total exposure time needed for the measurement while accounting for the natural variability of high-energy gamma ray sources, usually visible on the scale of days or weeks. This type of detector would not necessarily require an oriented tracker and/or calorimeter. Thus, it can be viewed as a completely separate solution with respect to the one discussed in the previous section.

As an example, we calculated the pair production cross section for 3~GeV fully linearly polarized photons incident on a cooled copper target, with the incidence glancing angle (i.e., the polar angle) close to one of the crystal main axes $\langle 101 \rangle$ ($\theta \approx 10$~mrad). In case of a polarization vector aligned along the (010) crystal plane (0$^\circ$) the cross section has a minimum (figure~\ref{fig:polar_xsec}), while in the opposite case (i.e., polarization orthogonal to the plane, 90$^\circ$ incidence) the cross section has a maximum.

\begin{figure*}[ht!]
    \centering
    \includegraphics[width=\linewidth]{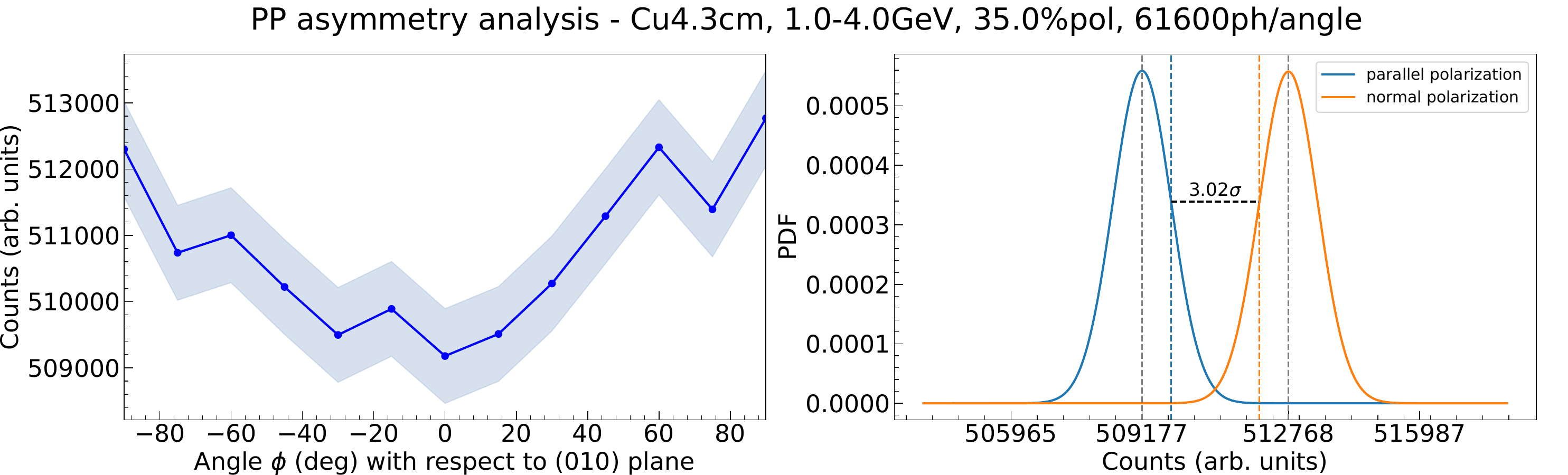}
    \caption{Left: Distribution of the number of pairs converting in the copper radiator as a function of the angle between the photon polarization vector and the vector normal to the crystallographic plane of the target. The dashed line represents the average value and the shaded area corresponds to $1\sigma$ from the mean. Right: Asymmetry analysis, which confirms the source polarization factor with a $3\sigma$ confidence level.}
    \label{fig:polar_asymmetry}
\end{figure*}

To investigate more quantitatively this idea, we have implemented in the Geant4 version 11.2 toolkit a custom model for the pair production class, where the cross section depends also on the relative angle between the crystal planes and the incident photon polarization. We have considered a simple setup, where a photon beam with energy uniformly distributed in the 1 -- 4~GeV range crosses a 4~m$^2$ large copper target foil with a variable thickness. We focused on the energy range where the maximum flux emitted by active galactic nuclei and pulsars is expected (\cite{ballet2024}). The transverse size of the target was chosen to be large, but still reasonably transportable by a modern launch vehicle. After the target, we have positioned a scoring grid, in order to score the number of {$\text{e}^+ - \text{e}^-$} pairs produced in the foil. We have repeated the simulation, changing each time the relative angle between the crystal plane and the polarization vector, in the $-90^\circ$ to $+90^\circ$ interval with a $15^\circ$ step, in order to reconstruct the full cross section variation. During the scan, the glancing angle ($\theta$) was fixed to 10~mrad. We have analyzed a total number of events corresponding to the observation for a little less than 6 consecutive months (15.05 Ms) of a source with a Vela pulsar-like source (i.e., PSR J0835-4510), assuming no dead time and no Earth occultation, a 35\% polarization fraction, and a flux of $1.33 \cdot 10^{-6}$~photons/(cm$^2$~s) between 1 and 4~GeV. This procedure was repeated for different target thicknesses, until the optimal thickness was found, namely the one which maximized the difference between the number of pairs produced in the normal polarization condition compared to the parallel polarization case. This difference is quantified by the number of sigma distances of the normal distributions of the counts that occur in the two conditions considered. The optimal target thickness was found to be 4.3~cm (3~X$_0$). As shown in figure~\ref{fig:polar_asymmetry}, we demonstrated that, under these hypotheses, it should be possible to identify the source polarization with a confidence level of $\sim$ 3 $\sigma$ ($<1\%$ false positive probability). Half of the time could be needed, if only a $90^\circ$ scan is performed, instead of a full $180^\circ$ one. Moreover, with higher polarization fractions, even smaller observation times could be achieved. It is also worth noting that the angle at which the maximum asymmetry between the normal and parallel polarization cross sections is expected ($\theta$) changes with energy, so considering an energy band without changing $\theta$ is more realistic and conservative than considering monochromatic photons.

This is already a demonstration that oriented crystals are a novel and viable tool for the measurement of the polarization of a steady $\gamma$-ray source. It should be noted that what we have presented are worst-case scenario estimates: the pair production cross section asymmetry (between different polarization states) is higher if electrons and positrons both carry around half of the photon energy (\cite{apyan2008}). Thus, by doing a more refined analysis with a more complex (position-sensitive) detector, where only a set of the acquired events is used, it should be possible to significantly lower the exposure necessary to acquire the desired statistics for a single source.  Moreover, by further optimizing the detector geometry (e.g., radiator thickness, material, or configuration), the data taking strategy (e.g., reducing the number of points in the angular scan), and the mission orbit (e.g., Lagrangian point orbit, where Earth occultation is minimal and Sun exposure is maximum), it should be feasible to achieve even better results. Finally, such a detector would surely achieve a large field of View, due to its small thickness and large effective area: multiple sources could thus be observed at the same time. In this sense, oriented crystals are definitely a viable solution which should be explored in the early future: the technology required to develop such detectors is already available, {as is} the one required to maintain the alignment of a satellite detector.

As a final note, it is worth mentioning that this technology enables polarization studies well beyond the few-gigaelectronvolts scale. In the 10--100~GeV range, the asymmetry can be significantly larger; despite the sharp decrease in photon flux, the required statistics is reduced by about an order of magnitude in this high-energy regime (\cite{apyan2008}).

\section{Technological challenges}
{It is worth briefly discussing what technological challenges will have to be addressed in the future, to allow the development of an oriented satellite detector. The first and most obvious challenge is how to grow oriented crystals at the scale required by a precision tracker or a calorimeter. As we have discussed also in section \ref{sec_oriented_sat}, state-of-the-art industrial techniques such as the Bridgman technique and the Czochralski method already allow the growth of a few crystalline materials, oriented along their axes and planes (\cite{senguttuvan1998}). Moreover, the OREO team has already demonstrated the feasibility of inter-aligning samples grown in different batches (\cite{malagutti2024,bandiera2023,bandiera2025}).}

A second set of challenges is related to the detectors coupled to the crystalline layers. On one hand, we do not foresee any issues in using the currently available silicon strip trackers and pixels along with oriented W or Cu crystals, both for the development of an oriented TKR (section~\ref{sec_oriented_tkr}) and for a polarization-sensitive detector (section~\ref{sec_polarization}). On the other hand, care should be used in choosing the right photodetectors for the readout of an oriented CAL. In fact, it was experimentally observed that the use of oriented crystalline scintillators increases the amount of light produced in a calorimeter bar up to a factor of three (\cite{soldani2025}). This means that high dynamic range detectors should be used, otherwise the saturation of the first few calorimeter layers could limit the high-energy detector performance. A possible solution is the employment of a scheme similar to the Fermi-LAT calorimeter, where both a low-gain and a high-gain photodiode are glued to the same crystalline bar (\cite{atwood2009,betti2024}).

A final set of challenges concerns the event reconstruction. In section \ref{sec_oriented_sat} we have shown that the response of an oriented detector changes significantly when passing from the on-axis to the off-axis alignment condition. This means that, to accurate process an event, it must be known if the incident photon is aligned with the crystal axis or not. The most accurate way to obtain this information is to measure the relative angle between the incident particle and the crystal axis. Thus, assuming that the spacecraft absolute position is measured with a precision at least as good as the Fermi-LAT ($\sim 2'' = 5.6 \cdot 10^{-4}$~deg), it is necessary to maintain the PSF below the SF angular scale ($\sim 0.1^\circ$) for energies above $\sim$ 10~GeV. This is already achievable with Fermi-LAT-like silicon detectors (\cite{atwood2009}), and can be further improved by employing techniques such as the floating strip readout scheme used in the AGILE silicon tracker (\cite{barbiellini2002}. A fine-grained measure of the shower development inside the calorimeter could also help in discriminating on-axis from off-axis particles, since the e.m. shower develops in different ways in the two cases (figure~\ref{fig:PWOcrys_Edep_vs_t}).

\section{Conclusions}\label{sec_conclusions}
In this paper we have demonstrated that oriented crystals can be used to tackle the most challenging aspects of satellite-based $\gamma$-ray astrophysics, essentially related to improving the angular resolution, the energy band-pass, the energy resolution, and the effective area, with respect to the layout of the Fermi Large Area Telescope, the instrument used as state-of-art reference for operating gamma ray space observatories.

Specifically, we have shown that high-energy photons would convert much earlier inside a silicon-tungsten tracker composed of oriented W crystals, and they would deposit a much larger fraction of their energy inside a scintillating calorimeter composed of oriented PbWO$_4$ bars. Considering the relative ease and cheapness required to grow these oriented crystals, two possible detector design strategies arise naturally:
\begin{enumerate}
    \item developing a standard $\gamma$-ray telescope employing oriented crystals instead of randomly aligned ones: such a detector would preserve its conventional response for off-axis incident particles, while achieving higher sensitivity and improved energy and angular resolution for on-axis photons;
    
    \item developing a thinner detector that performs comparably to a standard $\gamma$-ray telescope for on-axis photons, while being less sensitive for off-axis particles. The cost savings could be directed towards reducing the overall mission budget and weight, thus increasing the space mission feasibility and making it easier to rotate the spacecraft fast enough to observe short and intense electromagnetic transients such as GRBs. Payload compaction may also open the possibility to operate the instrumentation using nanosatellite platforms, (e.g., cubesats (\cite{cubesat})), largely improving the feasibility, schedule, and maximizing  launch opportunities through commercial deployers. Otherwise, it would be possible use the cost savings to maximize the satellite effective area without exceeding the initial budget, thus compensating the reduced sensitivity for off-axis photons and increasing even more the sensitivity for on-axis particles.
\end{enumerate}
Owing to the limited angular acceptance inherent to the strong field regime, such an oriented detector would be best optimized for pointing observations, targeted for instance to the Galactic Center or dwarf galaxies, to investigate the gigaelectronvolt gamma ray excess in the former and search in the latter for a high-energy $\gamma$-ray emission profile consistent with a dark matter signature.

The use of oriented crystals would also open up the possibility of measuring a $\gamma$-ray source polarization fraction at the few-gigaelectronvolt scale, thanks to the dependence of the pair production cross section on the $\gamma$-ray polarization, inside an oriented crystal. We demonstrated the feasibility of measuring a 35\% polarization fraction at the $3\sigma$ confidence level for a Vela pulsar-like source observed over the course of a few months. More complex and optimized layout could reduce the minimum detectable polarization fraction and/or the observation time. This will represent a real breakthrough in the field of $\gamma$-ray astrophysics, granting the possibility for substantial advances in the understanding of high-energy phenomena. At the present time, no other detection technique allows such a measurement, which would indeed be fundamental to confirm the nature of the Galactic Center $\gamma$-ray excess and the emission processes occurring inside astrophysical accelerators such as GRBs or supernova remnants.

The technology necessary to realize an oriented detector is already being developed, as demonstrated by the recent results obtained by the OREO collaboration, which at the present time is realizing and testing the first calorimeter composed by oriented crystals (\cite{bandiera2023, bandiera2025, malagutti2024}). The same group has also developed the Geant4 libraries used in this work to simulate numerically the interactions of high-energy particles inside oriented crystals (\cite{bandiera2015,sytov2019,sytov2023}). These studies are relevant not only for the future development of a properly oriented astroparticle detector, but also because they advance our general level of understanding of what happens inside $\gamma$-ray telescopes. Future steps foresee the development and test of a prototype tracker-calorimeter system, in order to better evaluate the improvements in energy and angular resolution that could be achieved thanks to the crystals orientation. Other tests fundamental for space-readiness will have to be performed (e.g., thermal, vibration and irradiation tests). Such a prototype detector could also be used as a platform to test the feasibility of measuring the gamma ray polarization at the gigaelectronvolt scale: it would be enough to replace the tungsten converters with an oriented copper radiator, and expose it to a polarized $\gamma$-ray beam produced with the same technique already used by the NA59 experiment at CERN (\cite{apyan2008}).

In conclusion, we believe that a detector composed of oriented crystals and employed in pointing strategy could be the answer to some of the currently open challenges in gamma ray astronomy and could help overcoming the limitations of direct gigaelectronvolt $\gamma$-ray observation, potentially revolutionizing the field in the post-Fermi era.

\appendix
\section*{Acknowledgements}
This work was primarily funded by INFN CSN5 through the OREO and Geant4-INFN projects.

\printbibliography

@article{senguttuvan1998,
    author = {N. Senguttuvan and others},
    title = {{Czochralski growth of lead tungstate single crystals and their characterization}},
    journal = {J. Cryst. Growth},
    year = {1998},
    volume = {183},
    number = {3},
    pages = {391-397},
    doi = {10.1016/S0022-0248(97)00437-5},
    issn = {0022-0248},
    
}

@article{uggerhoj2005,
	author = {Uggerh\o{}j, U.~I.},
    title = {{The interaction of relativistic particles with strong crystalline fields}},
	journal = {Rev. Mod. Phys.},
    year = {2005},
	volume = {77},
    number = {4},
	pages = {1131--1171},
	doi = {10.1103/RevModPhys.77.1131},
    publisher = {American Physical Society},
}

@article{sorensen1987_nature,
	author = {S\o{}rensen, A.~H. and Uggerh\o{}j, E.},
	title = {Channelling and channelling radiation},
	journal = {Nature},
    year = {1987},
	volume = {325},
    number = {6102},
    pages = {311-318},	
	doi = {10.1038/325311a0},
}

@article{baryshevski1989,
	author = {Baryshevsky, V.~G. and V.~V. Tikhomirov},
    title = {Synchrotron-type radiation processes in crystals and polarization phenomena accompanying them},
    journal = {Sov. Phys. Uspekhi},
    year = {1989},
    volume = {32},
    number = {11},
    pages = {1013},
    doi = {10.1070/PU1989v032n11ABEH002778},
}

@article{baryshevski1985,
	author = {Baryshevsky, V.~G. and V.~V. Tikhomirov},
    title = {Pair production in a slowly varying electromagnetic field and the pair production process},
    journal = {Phys. Lett. A},
	year = {1985},
    volume = {113},
	number = {6},
	pages = {335--340},
	doi = {10.1016/0375-9601(85)90178-1},
    issn = {0375--9601},	
}

@article{belkacem1987,
    author = {Belkacem, A. and others},
    title = {Study of $e^\pm$ pair creation by 20--150-GeV photons incident on a germanium crystal in alignment conditions},
    journal = {Phys. Rev. Lett.},
    year = {1987},
    volume = {58},
    number = {12},
    pages = {1196--1199},
    doi = {10.1103/PhysRevLett.58.1196},
    publisher = {American Physical Society},
}

@techreport{moller1989,
    author = "Moller, S. P. and others",
    title = "{CAN A CRYSTALLINE DETECTOR BE USED IN VERY HIGH-ENERGY GAMMA-RAY ASTRONOMY?}",
    reportNumber = "CERN-EP-89-149",
    year = "1989",
    url = "https://cds.cern.ch/record/202865",
    institution = "CERN"
}

@article{kirsebom1998,
	author = {Kirsebom, K. and others},
    title = {{Pair production by 5--150 GeV photons in the strong crystalline fields of germanium, tungsten and iridium}},
	journal = {Nucl. Instrum. Methods Phys. Res. B},
	year = {1998},
    volume = {135},
	number = {1},
	pages = {143--148},
	doi = {10.1016/S0168-583X(97)00589-2},
    issn = {0168-583X},
}

@article{moore1996,
	author = {Moore, R. and others},
    title = {{Measurement of pair-production by high energy photons in an aligned tungsten crystal}},
	journal = {Nucl. Instrum. Methods Phys. Res. B},
	year = {1996},
    volume = {119},
	number = {1},
	pages = {149--155},
	doi = {10.1016/0168-583X(96)00347-3},
    issn = {0168-583X},	
}

@article{baskov1999,
	author = {Baskov, V.~A. and others},
    title = {Electromagnetic cascades in oriented crystals of garnet and tungstate},
	journal = {Phys. Lett. B},
    year = {1999},
	volume = {456},
	number = {1},
	pages = {86--89},
	doi = {10.1016/S0370-2693(99)00444-X},
	issn = {0370-2693},	
}

@article{cabibbo1962,
    author = {Cabibbo, N. and others},
    title = {New Method for Producing and Analyzing Linearly Polarized Gamma-Ray Beams},
    journal = {Phys. Rev. Lett.},
    year = {1962},
    volume = {9},
    number = {6},
    pages = {270--272},
    doi = {10.1103/PhysRevLett.9.270},
    publisher = {American Physical Society},
}

@article{cabibbo1962circular,
    author = {Cabibbo, N. and others},
    title = {Circular polarization of high-energy $\gamma$ rays by birefringence in crystals},
    journal = {Phys. Rev. Lett.},
    year = {1962},
    volume = {9},
    number = {10},
    pages = {435--437},
    doi = {10.1103/PhysRevLett.9.435},
    publisher = {American Physical Society},
}

@article{palazzi1968high,
  title={High-energy bremsstrahlung and electron pair production in thin crystals},
  author={Palazzi, Giordano Diambrini},
  journal={Reviews of Modern Physics},
  volume={40},
  number={3},
  pages={611},
  year={1968},
  publisher={APS}
}

@article{apyan2008,
    author = {Apyan, A. and {NA59 Collaboration}},
    title = {{Coherent bremsstrahlung, coherent pair production, birefringence, and polarimetry in the 20--170 GeV energy range using aligned crystals}},
    journal = {Phys. Rev. ST Accel. Beams},
    year = {2008},
    volume = {11},
    number = {4},
    pages = {041001},  
    publisher = {American Physical Society},
    doi = {10.1103/PhysRevSTAB.11.041001},
}

@book{baier1998_book,
    author = {Baier, V.~N. and Katkov, V.~M. and Strakhovenko, V.~M.},
    title = {Electromagnetic Processes at High Energies in Oriented Single Crystals},
    year = {1998},
    doi = {10.1142/2216},
    publisher = "World Scientific",
    address = {Singapore},
}

@article{chehab2002,
    author = {Chehab, R. and others},
    title = {Experimental study of a crystal positron source},
    journal = {Phys. Lett. B},
    year = {2002},
    volume = {525},
    number = {1},
    pages = {41--48},
    doi = {10.1016/S0370-2693(01)01395-8},
    issn = {0370-2693},
}

@article{artru2005,
    author = {Artru, X. and others},
    title = {{Summary of experimental studies, at CERN, on a positron source using crystal effects}},
    journal = {Nucl. Instrum. Methods Phys. Res. B},
    year = {2005},
    volume = {240},
    number = {3},
    pages = {762--776},
    doi = {10.1016/j.nimb.2005.04.134},
    issn = {0168-583X},    
}

@article{diambrinipalazzi1962,
	author = {Diambrini-Palazzi, G.},
    title = {Interazioni di fotoni ed elettroni di alta energia in cristalli},
    journal = {Nuovo Cim.},
    year = {1962},
	volume = {25},
	number = {2},
	pages = {88--99},
    doi = {10.1007/BF02860173},
}

@article{bandiera2025,
    author = {Bandiera, L. and Fedeli, P. and {The Oriented Calorimeter Collaboration}},
    title={{High-performance electromagnetic calorimeter with oriented crystals to open new pathways in particle and astroparticle physics}},
    year = {2025},
    journal = {Front. Sens.},
    volume = {6},
    number = {2025},
    pages = {},
    doi = {10.3389/fsens.2025.1659893},
    issn = {2673-5067},
}

@article{bandiera2018,
	author = {Bandiera, L. and others},
    title = {Strong Reduction of the Effective Radiation Length in an Axially Oriented Scintillator Crystal},
	journal = {Phys. Rev. Lett.},
	year = {2018},
    volume = {121},
	number = {2},
	pages = {021603},
	doi = {10.1103/PhysRevLett.121.021603},
    publisher = {American Physical Society},
}

@article{bandiera2023,
	author = {Bandiera, L. and others},
    title = {A highly-compact and ultra-fast homogeneous electromagnetic calorimeter based on oriented lead tungstate crystals},
	journal = {Front. Phys.},
    year = {2023},
	volume = {11},
	number = {1254020},
	pages = {1--11},
	doi = {10.3389/fphy.2023.1254020},
}

@article{soldani2025,
    author={Soldani, M. and Monti-Guarnieri, P. and Selmi, A. and others},
    title = {{Strong enhancement of electromagnetic shower development in oriented scintillating crystals and implications for particle detectors}}, 
    journal = {Eur. Phys. J. C},
    year={2025},
    volume = {85},
    number = {1239},
    pages = {1--10},
    doi={10.1140/epjc/s10052-025-14967-4}
}

@misc{soldani2023phd,
  author = {Soldani, M.},
  title = {{Innovative applications of strong crystalline field effects to particle accelerators and detectors}},
  howpublished = {{PhD thesis, University of Ferrara}, \url{https://repository.cern/records/qan11-jzm91}},
  year = {2023},
}

@article{bandiera2019,
    author = {L. Bandiera and others},
    title = {Compact electromagnetic calorimeters based on oriented scintillator crystals},
    journal = {Nucl. Inst. Methods Phys. Res. A},
    year = {2019},
    volume = {936},
    issue = "{Frontier Detectors for Frontier Physics: 14th Pisa Meeting on Advanced Detectors}",
    pages = {124--126},
    doi = {10.1016/j.nima.2018.07.085},
    issn = {0168-9002},
}

@article{sytov2023,
    author = {Sytov, A. and others},
    title = {{Geant4 simulation model of electromagnetic processes in oriented crystals for accelerator physics}},
	journal = {J. Korean Phys. Soc.},
	year = {2023},
    volume = {83},
    issue = {{Special Issue on the 24th International Conference on Accelerators and Beam Utilizations (ICABU 2022) I}},
	pages = {132--139},
	doi = {10.1007/s40042-023-00834-6},
}

@article{guidi2012,
	author = {Guidi, V. and others},
    title = {Radiation generated by single and multiple volume reflection of ultrarelativistic electrons and positrons in bent crystals},
	journal = {Phys. Rev. A},
    year = {2012},
	volume = {86},
	issue = {4},
	pages = {042903},
	doi = {10.1103/PhysRevA.86.042903},
    publisher = {American Physical Society},
}

@article{bandiera2015,
	author = {Bandiera, L. and others},
    title = {{RADCHARM++: A C++ routine to compute the electromagnetic radiation generated by relativistic charged particles in crystals and complex structures}},
	journal = {Nucl. Instrum. Methods Phys. Res. B},
	year = {2015},
    volume = {355},
    issue = {{Proceedings of the 6th International Conference Channeling 2014}},
	pages = {44--48},
	doi = {10.1016/j.nimb.2015.03.031},
    issn = {0168-583X},	
}

@article{sytov2019,
	author = {Sytov, A. and others},
    title = {Simulation code for modeling of coherent effects of radiation generation in oriented crystals},
	journal = {Phys. Rev. Accel. Beams},
    year = {2019},
	volume = {22},
	number = {6},
	pages = {064601},
	doi = {10.1103/PhysRevAccelBeams.22.064601},
    publisher = {American Physical Society},
}

@article{baryshevsky2017,
	author = {Baryshevsky, V.~G. and others},
    title = {On the influence of crystal structure on the electromagnetic shower development in the lead tungstate crystals},
	journal = {Nucl. Instrum. Methods Phys. Res. B},
    year = {2017},
	volume = {402},
    issue = {Proceedings of the 7th International Conference Channeling 2016},
	pages = {35--39},
	doi = {10.1016/j.nimb.2017.02.066},
    issn = {0168-583X},	
}

@article{soldani2023,
  author = {Soldani, M. and others},
  title = {Strong enhancement of electromagnetic shower development induced by high-energy photons in a thick oriented tungsten crystal},
  journal = {Eur. Phys. J. C},
  year = {2023},
  volume = {83},
  number = {101},
  pages = {1--8},
  doi = {10.1140/epjc/s10052-023-11247-x},
}

@article{montiguarnieri2024,
    author = {Monti-Guarnieri, P. and others},
    title = {Particle identification capability of a homogeneous calorimeter composed of oriented crystals},
    journal = {JINST},
    year = {2024},
    volume = {19},
    number = {10},
    pages = {P10014},
    doi = {10.1088/1748-0221/19/10/P10014},
    publisher = {IOP Publishing},
}

@article{bandiera2022,
	author = {Bandiera, L. and others},
    title = {Crystal-based pair production for a lepton collider positron source},
    journal = {Eur. Phys. J. C},
    year = {2022},
    volume = {82},
    number = {699},
    pages = {1--13},
    doi = {10.1140/epjc/s10052-022-10666-6},
}

@article{malagutti2024,
    author = {Malagutti, L. and others},
    title = {High-precision alignment techniques for realizing an ultracompact electromagnetic calorimeters using oriented high-Z scintillator crystals},
    journal = {Nucl. Instrum. Methods Phys. Res. A},
    year = {2024},
    volume = {1069},
    issue = "{Frontier Detectors for Frontier Physics: 16th Pisa Meeting on Advanced Detectors}",
    pages = {169869},
    doi = {10.1016/j.nima.2024.169869},
    issn = {0168-9002},
}

@article{atwood2009,
    author = {Atwood, W.~B. and others},
    title = {{The Large Area Telescope on the FERMI gamma-ray space telescope mission}},
    journal = {{ApJ}},
    year = {2009},
    volume = {697},
    number = {2},
    pages = {1071},	
    doi = {10.1088/0004-637X/697/2/1071},
	publisher = {The American Astronomical Society},
}

@article{ackermann2012,
	author = {Ackermann, M. and others},
    title = {{The Fermi Large Area Telescope on Orbit: Event Classification, Instrument Response Functions, and Calibrations}},
    journal = {{ApJ}},
    year = {2012},
    volume = {203},
    number = {1},
    pages = {4},
    doi = {10.1088/0067-0049/203/1/4},
	publisher = {The American Astronomical Society},    
}

@misc{ATel2024_16450,
    author = {Tavani, M. and others},
    title = "{The AGILE satellite ceased operations and re-entered today into the atmosphere}",
    howpublished = {ATel 16450, \url{https://www.astronomerstelegram.org/?read=16450}},
    year = 2024,
}

@article{tavani2009,
	author = {Tavani, M. and others},
	title = {{The AGILE Mission}},
	journal = {A\&A},
    year = 2009,
    volume = 502,
    number = 3,
    pages = "995-1013",
    doi = "10.1051/0004-6361/200810527",
}

@article{atwood2013,
    author={Atwood, W.~B.and others},
    title={{Pass 8: Toward the Full Realization of the Fermi-LAT Scientific Potential}}, 
    journal = {ArXiv e-prints},
    year={2013},
    eprint={1303.3514},
    archivePrefix={arXiv},
    primaryClass={astro-ph.IM},
    doi={10.48550/arXiv.1303.3514},
}

@article{ballet2024,
      author={Ballet, J. and others},
      title={Fermi Large Area Telescope Fourth Source Catalog Data Release 4 (4FGL-DR4)}, 
      year={2024},
      journal = {ArXiv e-prints},
      eprint={2307.12546},
      archivePrefix={arXiv},
      primaryClass={astro-ph.HE},
      doi={10.48550/arXiv.2307.12546}, 
}

@article{fermilatDR3,
	author = {Abdollahi, S. and others},
	title = {{Incremental Fermi Large Area Telescope Fourth Source Catalog}},
	journal = {ApJS},
    year = 2022,
    volume = 260,
    number = 53,
    pages = "1--24",
    doi = "10.3847/1538-4365/ac6751", 
}

@book{aharonian2004,
    author = {Aharonian, F.~A.},
    title = {Very high energy cosmic gamma radiation : a crucial window on the extreme Universe},
    year = {2004},
    doi = {10.1142/4657},
    publisher = "World Scientific",
    address = {Singapore},
}

@article{atwood2007,
    author = {Atwood, W.~B. and others},
    title = {Design and initial tests of the Tracker-converter of the Gamma-ray Large Area Space Telescope},
    journal = {Astropart. Phys.},
    year = {2007},
    volume = {28},
    number = {4},
    pages = {422-434},
    doi = {10.1016/j.astropartphys.2007.08.010},
    issn = {0927-6505},
}

@article{barbiellini2002,
    author = {Barbiellini, G. and others},
    title = {{The AGILE silicon tracker: testbeam results of the prototype silicon detector}},
    journal = {Nucl. Instr. Methods Phys. Res. A},
    year = {2002},
    volume = {490},
    number = {1},
    pages = {146-158},
    doi = {10.1016/S0168-9002(02)01062-8},
    issn = {0168-9002},
}

@article{bartels2016,
    author = {Bartels, R. and others},
    title = {{Strong Support for the Millisecond Pulsar Origin of the Galactic Center GeV Excess}},
    journal = {Phys. Rev. Lett.},
    year = {2016},
    volume = {116},
    number = {5},
    pages = {051102},   
    doi = {10.1103/PhysRevLett.116.051102},
    publisher = {American Physical Society},
}

@article{charles2014,
    author={Charles, Eric},
    title="{Scientific motivations and technical design considerations for future high-energy gamma-ray telescopes in light of lessons learned from the Fermi Large Area Telescope}",
    journal={Space Telescopes and Instrumentation 2014: Ultraviolet to Gamma Ray},
    year={2014},
    volume={9144},
    number={1},
    pages={100--119},
    doi={10.1117/12.2057018},
    publisher={SPIE},
}

@article{ilie2019,
    author = {C. Ilie},
    title = {{Gamma-Ray Polarimetry: A New Window for the Nonthermal Universe}},
    journal = {PASP},
    year = {2019},
    volume = {131},
    number = {1005},
    pages = {111001},
    doi = {10.1088/1538-3873/ab2a3a},
    publisher = {The Astronomical Society of the Pacific},
}

@article{bernard2013,
    author = {D. Bernard},
    title = {{HARPO -- A gaseous TPC for high angular resolution $\gamma$-ray astronomy and polarimetry from the MeV to the TeV}},
    journal = {Nucl. Instr. Methods Phys. Res. A},
    year = {2013},
    volume = {718},
    issue = {Proceedings of the 12th Pisa Meeting on Advanced Detectors},
    pages = {395-399},
    doi = {10.1016/j.nima.2012.10.054},
    issn = {0168-9002},
}

@article{takahashi2015,
    author = {Takahashi, S. and others},
    title = "{GRAINE project: The first balloon-borne, emulsion gamma-ray telescope experiment}",
    journal = {Progr. Theor. Exp.},
    year = {2015},
    volume = {2015},
    number = {4},
    pages = {043H01},
    doi = {10.1093/ptep/ptv046},
    issn = {2050-3911},
}

@inbook{bernard2024,
    author = "Bernard, D. and others",
    title = "Gamma-ray polarimetry",
    bookTitle = "Handbook of X-ray and Gamma-ray Astrophysics",
    year = "2022",
    publisher = "Springer Nature Singapore",
    address = "Singapore",
    isbn = "978-981-16-4544-0",
    doi = "10.1007/978-981-16-4544-0_52-1",
    pages = {1-44},
}

@inbook{delmonte2022,
    author = "Del Monte, E. and others",
    title = "Compton Polarimetry",
    bookTitle = "Handbook of X-ray and Gamma-ray Astrophysics",
    year = "2022",
    publisher = "Springer Nature Singapore",
    address = "Singapore",
    isbn = "978-981-16-4544-0",
    doi = "110.1007/978-981-16-4544-0_27-1",
    pages = {1-45},
}

@article{soffitta2024,
    author = {Soffitta, P.},
    title = {{The Imaging X-ray Polarimetry Explorer (IXPE) and New Directions for the Future}},
    journal = {Instruments},
    year = {2024},
    volume = {8},
    number = {2},
    pages = {25},
    doi = {10.3390/instruments8020025},
    issn = {2410-390X},
}

@article{bernard2022,
    author = {D. Bernard},
    title = {{MeV$-$GeV polarimetry with $\gamma \to e^+ e^-$: Asserting the performance of silicon strip detectors-based telescopes}},
    journal = {Nucl. Instrum. Methods Phys. Res. A},
    year = {2022},
    volume = {1042},
    number = {},
    pages = {167462},
    doi = {10.1016/j.nima.2022.167462},
    issn = {0168-9002},    
}

@article{bringmann2012,
    author = {T. Bringmann and C. Weniger},
    title = {{Gamma ray signals from dark matter: Concepts, status and prospects}},
    journal = {Phys. Dark Universe},
    year = {2012},
    volume = {1},
    number = {1},
    pages = {194-217},
    doi = {10.1016/j.dark.2012.10.005},
    issn = {2212-6864},  
}

@article{leane2019,
    author = {Leane, R.~K. and Slatyer, T.~R.},
    title = {{Revival of the Dark Matter Hypothesis for the Galactic Center Gamma-Ray Excess}},
    journal = {Phys. Rev. Lett.},
    year = {2019},
    volume = {123},
    issue = {24},
    pages = {241101},
    doi = {10.1103/PhysRevLett.123.241101},
    publisher = {American Physical Society},
}

@article{hoffman1999,
    author = {Hoffman, C. M. and others},
    title = {{Gamma-ray astronomy at high energies}},
    journal = {Rev. Mod. Phys.},
    year = {1999},
    volume = {71},
    issue = {4},
    pages = {897--936},
    doi = {10.1103/RevModPhys.71.897},
    publisher = {American Physical Society},
}

@article{funk2015,
    author = "Funk, S.",
    title = "Ground- and Space-Based Gamma-Ray Astronomy", 
    journal= "Annu. Rev. Nucl",
    year = "2015",
    volume = "65",
    number = {},
    pages = "245--277",
    doi = "10.1146/annurev-nucl-102014-022036",
    publisher = "Annual Reviews",
    issn = "1545-4134",
  }

@article{fleischhack2021,
    author = "Fleischhack, H.",
    title = "{AMEGO-X: MeV gamma-ray Astronomy in the Multi-messenger Era}",
    journal = "{PoS}",
    year = 2021,
    volume = "395",
    issue = "{Proceedings of 37th International Cosmic Ray Conference -- PoS(ICRC2021)}",
    pages = "649",
    doi = "10.22323/1.395.0649",
}

@article{deangelis2021,
    author = {De Angelis, A. and others},
    title = {{Gamma-ray astrophysics in the MeV range}},
    journal = {Exp. Astron.},
    year = {2021},
    volume = {51},
    number = {},
    pages = {1225--1254},
    doi = {10.1007/s10686-021-09706-y}
}

@article{gueta2021,
    author = "Gueta, Orel",
    title = "{{The Cherenkov Telescope Array: layout, design and performance}}",
    journal = "{PoS}",
    year = 2021,
    volume = "395",
    issue = "{Proceedings of 37th International Cosmic Ray Conference -- PoS(ICRC2021)}",
    pages = "885",
    doi = "10.22323/1.395.0885",
}

@article{abdollahi2024,
      title={{Search for Extended GeV Sources in the Inner Galactic Plane}}, 
      author={S. Abdollahi and others},
      year={2024},
      journal = {ArXiv e-prints},
      eprint={2411.07162},
      archivePrefix={arXiv},
      primaryClass={astro-ph.HE},
      doi = {10.48550/arXiv.2411.07162}
}

@article{torii2007,
    author = {S. Torii and {The CALET collaboration}},
    title = {{The CALET experiment on ISS}},
    journal = {Nucl. Phys. B},
    year = {2007},
    volume = {166},
    issue = "{Proceedings of the Third International Conference on Particle and Fundamental Physics in Space}",
    pages = {43-49},
    doi = {10.1016/j.nuclphysbps.2006.12.046},
    issn = {0920-5632},
}

@article{kai2025,
    author = {Kai-Kai, D. and others},
    title = {{PSF calibration of DAMPE for gamma-ray observations}},
    journal = {Astropart. Phys.},
    year = {2025},
    volume = {165},
    number = {},
    pages = {103058},
    doi = {10.1016/j.astropartphys.2024.103058},
    url = {https://www.sciencedirect.com/science/article/abs/pii/S092765052400135X},
    issn = {0927-6505},
}

@article{kyratzis2022,
    author = {D. Kyratzis and others},
    title = {{Overview of the HERD space mission}},
    journal = {Phys. Scr.},
    year = {2022},
    volume = {97},
    number = {5},
    pages = {054010},
    doi = {10.1088/1402-4896/ac63fc},
}

@article{totani2025,
    author = {Totani, Tomonori},
    title = {{20 GeV halo-like excess of the Galactic diffuse emission and implications for dark matter annihilation}},
    journal = {JCAP},
    year = {2025},
    volume = {2025},
    number = {11},
    pages = {080},
    doi = {10.1088/1475-7516/2025/11/080},
    publisher = {IOP Publishing},
}

@article{betti2024,
    author = {Betti, P. and others},
    title = {{Double Photodiode Readout System for the Calorimeter of the HERD Experiment: Challenges and New Horizons in Technology for the Direct Detection of High-Energy Cosmic Rays}},
    journal = {Instruments},
    year = {2024},
    volume = {8},
    issue = {1},
    pages = {5},
    doi = {10.3390/instruments8010005},
}

@article{allison2016,
	author = {Allison, J. and others},
    title = {{Recent developments in Geant4}},
	journal = {Nucl. Instrum. Methods Phys. Res. A},
    year = {2016},
	volume = {835},
    number = {},
	pages = {186--225},
	doi = {10.1016/j.nima.2016.06.125},
    issn = {0168-9002},	
}

@article{agostinelli2003,
    author = {S. Agostinelli and others},
    title = {Geant4 -- a simulation toolkit},
    journal = {Nucl. Inst. Methods Phys. Res. A},
    year = {2003},
    volume = {506},
    number = {3},
    pages = {250-303},
    doi = {10.1016/S0168-9002(03)01368-8},
    issn = {0168-9002},    
}

@article{fabjan2003,
	author = {Fabjan, C.~W. and Gianotti, F.},
    title = {Calorimetry for particle physics},
	journal = {Rev. Mod. Phys.},
	year = {2003},
    volume = {75},
	issue = {4},
    pages = {1243},
	doi = {10.1103/RevModPhys.75.1243},
    publisher = {American Physical Society},
}

@article{Dinsmore:2021nip,
    author = "Dinsmore, Jack T. and Slatyer, Tracy R.",
    title = "{Luminosity functions consistent with a pulsar-dominated Galactic Center excess}",
    journal = "JCAP",
    year = "2022",
    volume = "06",
    number = "06",
    pages = "025",
    doi = "10.1088/1475-7516/2022/06/025",
}

@article{negrello2025novel,
    author={Negrello, R. and others},
    title={{A novel tool for advanced analysis of Geant4 simulations of charged particles interactions in oriented crystals}},
    journal = {Nucl. Instrum. Methods Phys. Res. A},
    year={2025},
    volume={1074},
    issue = "{Charged and Neutral Particles Channeling Phenomena - Channeling 2024}",
    pages={170277},
    doi = {10.1016/j.nima.2025.170277},
    publisher={Elsevier},
}

@techreport{cubesat,
    author      = {{The CubeSat Program}},
    title       = {{CubeSat Design Specification (1U - 12U)}},
    institution = {{Cal Poly CubeSat Laboratory}},
    year        = {{2022}},
    number      = {{CP-CDS-R14.1}},
    note        = {{REV 14.1}}
}
\end{document}